\documentclass[10pt,journal,compsoc, onecolumn]{IEEEtran}

\usepackage{cite}
\usepackage{amsmath}
\usepackage{algorithmic}
\usepackage{array}
\usepackage[caption=false,font=footnotesize,labelfont=sf,textfont=sf]{subfig}
\usepackage{url}
\usepackage{graphicx}
\usepackage{multirow}
\usepackage{color}

\begin{document}
%
\title{Architectures and Key Technical Challenges for 5G Systems Incorporating Satellites}
%
%
%
%

\author{A.~Guidotti,~\IEEEmembership{Member,~IEEE,}
        A.~Vanelli-Coralli,~\IEEEmembership{Senior Member,~IEEE,}
        M.~Conti,
        S.~Andrenacci,~\IEEEmembership{Member,~IEEE,}
        S.~Chatzinotas,~\IEEEmembership{Senior Member,~IEEE}
        N.~Maturo,
        B.~Evans,~\IEEEmembership{Senior Member,~IEEE}
        A.~Awoseyila,~\IEEEmembership{Member,~IEEE}
        A.~Ugolini, T.~Foggi, L.~Gaudio,
        N.~Alagha,~\IEEEmembership{Member,~IEEE}
        S.~Cioni,~\IEEEmembership{Senior Member,~IEEE}
\IEEEcompsocitemizethanks{\IEEEcompsocthanksitem Alessandro Guidotti is the corresponding author. E-mail: a.guidotti@unibo.it \protect\\
\IEEEcompsocthanksitem
Alessandro Guidotti, Alessandro Vanelli-Coralli, and Matteo Conti are with the Department of Electrical, Electronic, and Information Engineering (DEI) ``Guglielmo Marconi,'' Univ. of Bologna, Italy. \protect\\
Stefano Andrenacci, Symeon Chatzinotas, and Nicola Maturo are with the Interdisciplinary Centre for Security, Reliability and Trust (SnT), Univ. of Luxembourg, Luxembourg. \protect\\
Barry Evans and Adegbenga Awoseyila are with the Institute for Communication Systems (ICS), University of Surrey, Guildford, UK.  \protect\\
Alessandro Ugolini and Lorenzo Gaudio are  with Department of Engineering and Architecture, Univ. of Parma, Italy. \protect\\
Tommaso Foggi is with CNIT Research Unit, Department of Engineering and Architecture, Univ. of Parma, Italy.\protect\\
Nader Alagha and Stefano Cioni are with European Space Agency - esa.int, ESTEC/TEC-EST, Noordwijk, The Netherlands.}
}

%

\IEEEtitleabstractindextext{%
\begin{abstract}
Satellite Communication systems are a promising solution to extend and complement terrestrial networks in unserved or under-served areas. This aspect is reflected by recent commercial and standardisation endeavours. In particular, 3GPP recently initiated a Study Item for New Radio-based, \emph{i.e.}, 5G, Non-Terrestrial Networks aimed at deploying satellite systems either as a stand-alone solution or as an integration to terrestrial networks in mobile broadband and machine-type communication scenarios. However, typical satellite channel impairments, as large path losses, delays, and Doppler shifts, pose severe challenges to the realisation of a satellite-based NR network. In this paper, based on the architecture options currently being discussed in the standardisation fora, we discuss and assess the impact of the satellite channel characteristics on the physical and Medium Access Control layers, both in terms of transmitted waveforms and procedures for enhanced Mobile BroadBand (eMBB) and NarrowBand-Internet of Things (NB-IoT) applications. The proposed analysis shows that the main technical challenges are related to the PHY/MAC procedures, in particular Random Access (RA), Timing Advance (TA), and Hybrid Automatic Repeat reQuest (HARQ) and, depending on the considered service and architecture, different solutions are proposed.
\end{abstract}
}

\maketitle

\IEEEdisplaynontitleabstractindextext

 \ifCLASSOPTIONpeerreview
 \begin{center} \bfseries EDICS Category: 3-BBND \end{center}
 \fi
%
\IEEEpeerreviewmaketitle

\IEEEraisesectionheading{\section{Introduction}\label{sec:introduction}}

\IEEEPARstart{A}{n} ever growing demand for broadband high-speed, heterogeneous, ultra-reliable, secure, and low latency services recently started being experienced in wireless communications. These drivers require enhancements to devices, services, and technologies that are currently well established in the global market, as for instance the 3GPP Long Term Evolution standard. Thus, the definition of new standards and technologies, known as 5G, has become of outmost importance in order to introduce novel techniques and technologies that can support the fulfillment of the significantly demanding requirements as well as to support novel market segments. The massive scientific and industrial interest in 5G communications is in particular motivated by the key role that these future system will play in the global economic and societal processes to support the next generation vertical services, \emph{e.g.}, Internet of Things, automotive and transportation sectors, e-Health, Industry 4.0, etc., \cite{5GPPP_1,5GPPP_2}.

Unlike previous standards, which can be seen as general-purpose technologies to which the different services were tailored and adjusted, the next 5G standard is expected to be able to provide tailored and optimised support for a plethora of services, traffic loads, and end-user communities. Such a technology revolution can only be achieved by means of a radical shift in the way both the access and the core network are designed. This heterogeneous and optimised framework is reflected in the challenging requirements that 5G systems are expected to meet, \emph{e.g.}, large throughput increase (the target peak data rate should be $20$ Gbps in the downlink and $10$ Gbps in the uplink), global and seamless connectivity, reliability ($99.999\%$ of successful packet reception), and connection density ($1$ million devices per square km), amongst others, \cite{3GPP_38913}. In this context, the integration of satellite and terrestrial networks can be a cornerstone to the realisation of the foreseen heterogeneous global system. Thanks to their inherently large footprint, satellites can efficiently complement and extend dense terrestrial networks, both in densely populated areas and in rural zones, as well as provide reliable Mission Critical services. The definition of the new 5G paradigm provides a unique opportunity for the terrestrial and satellite communities to define an harmonised and fully-fledged architecture, differently from the past when terrestrial and satellite networks evolved almost independently from each other, leading to a difficult \emph{a posteriori} integration. This trend is substantiated by 3GPP Radio Access Network (RAN) and Service and system Aspects (SA) activities, in which a new Study Item has recently started on Non-Terrestrial Networks (NTN) for 5G systems, \cite{3GPP_SatSI,3GPP_38811}. The role of NTN in 5G systems is expected to be manifold, including: i) the support to 5G service provision in both un-served areas that cannot be covered by terrestrial 5G networks (isolated/remote areas, on board aircrafts or vessels) and underserved areas (\emph{e.g.}, sub-urban/rural areas); ii) improve the 5G service reliability thanks to a better service continuity, in particular for mission critical communications or Machine Type Communications (MTC) and Internet of Things (IoT) devices, M2M/IoT devices or for passengers on board moving platforms; and iii) to enable the 5G network scalability by providing efficient multicast/broadcast resources for data delivery. In addition to the 3GPP standardisation effort, which will be further detailed in this paper, also funded projects are currently addressing SatCom-based 5G systems, as, for instance: i) the EC H2020 project VITAL (VIrtualized hybrid satellite-Terrestrial systems for resilient and fLexible future networks), in which the combination of terrestrial and satellite networks is addressed by bringing Network Functions Virtualization (NFV) into the satellite domain and by enabling Software-Defined-Networking (SDN)-based, federated resources management in hybrid SatCom-terrestrial networks, \cite{VITAL}; ii) the EC H2020 project Sat5G (Satellite and Terrestrial Network for 5G), which aims at developing a cost-effective SatCom solution by means of satellite gateway and terminal virtualisation and use cases demonstration \cite{Sat5G}; iii) the ESA project SATis5 (Demonstrator for Satellite Terrestrial Integration in the 5G Context), in which a set of relevant satellite use cases for 5G in the areas of enhanced Mobile BroadBand (eMBB) and massive IoT deployments will be demonstrated, \cite{Satis5G}; and iv) the H2020 project SANSA (Shared Access Terrestrial-Satellite Backhaul Network enabled by Smart Antennas), aimed at to enhancing the performance of mobile wireless backhaul networks, both in terms of capacity and resilience, while assuring an efficient use of the spectrum, \cite{SANSA}.

In the above context for worldwide 5G systems, the integration of terrestrial systems with Geostationary Earth Orbit (GEO) satellites would be beneficial for global large-capacity coverage, but the large delays in geostationary orbits pose significant challenges, as will be also highlighted in this work. In \cite{Intro1,Intro2,Intro3}, resource allocation algorithms for multicast transmissions and TCP protocol performance were analysed in a Long Term Evolution (LTE)-based GEO system, providing valuable solutions. However, to avoid the above issues, significant attention is being gained by Low Earth Orbit (LEO) \emph{mega-constellations}, \emph{i.e.}, systems in which hundreds of satellites are deployed to provide global coverage, as also demonstrated by recent commercial endeavours.  In \cite{Intro_UniBo,Intro_UniBo2}, a mega-constellation of LEO satellites deployed in Ku-band to provide LTE broadband services was proposed and the impact of typical satellite channel impairments as large Doppler shifts and delays was assessed with respect to the PHY and MAC layer procedures. The introduction of SatCom in 5G systems has been preliminarily addressed by the authors in \cite{Intro_UniBo3,Intro_UniBo4}, in which the focus was on the PHY and MAC layer techniques that shall cope with satellite channel impairments. However, these works are based on several assumptions due to the fact that the 3GPP standardisation for SatCom-based 5G was still in its infancy.
3GPP studies and activities are now providing significant results and critical decisions have been made on the PHY and MAC layers for the New Radio (NR) air interface. In particular, the first PHY standard has been made available and preliminary analyses related to the deployment of 5G systems through SatCom are advancing. In this context, it is of outmost importance to assess the impact that these new requirements will have on future 5G Satellite Communications (SatCom). To this aim, in this paper, we move from the analysis performed in \cite{Intro_UniBo,Intro_UniBo2,Intro_UniBo3,Intro_UniBo4} and assess the impact of large delays and Doppler shifts in two scenarios of interest for future 5G systems, one for enhanced Mobile BroadBand (eMBB) services and one for NarrowBand-IoT (NB-IoT), for which significantly different system architectures are considered.

The remainder of this paper is organised as follows. In Section~\ref{sec:Architecture}, we outline the overall system architecture for both the enhanced Mobile BroadBand and NarrowBand-IoT scenarios and introduce the main satellite channel impairments that are considered in the following analyses. In Section~\ref{sec:eMBB}, the waveform and main PHY/MAC procedures for NR are discussed and the impact of the previously introduced channel impairments is discussed. In Section~\ref{sec:NBIOT}, the main characteristics of NB-IoT communications are outlined with respect to the considered channel impairments and preliminary numerical results are provided. Finally, Section~\ref{sec:Conclusions} concludes this work.

\section{System Architecture}
\label{sec:Architecture}
\begin{figure}[t!]
     \subfloat[A1.\label{fig:ArchitectureA1}]{%
       \includegraphics[width=0.5\textwidth]{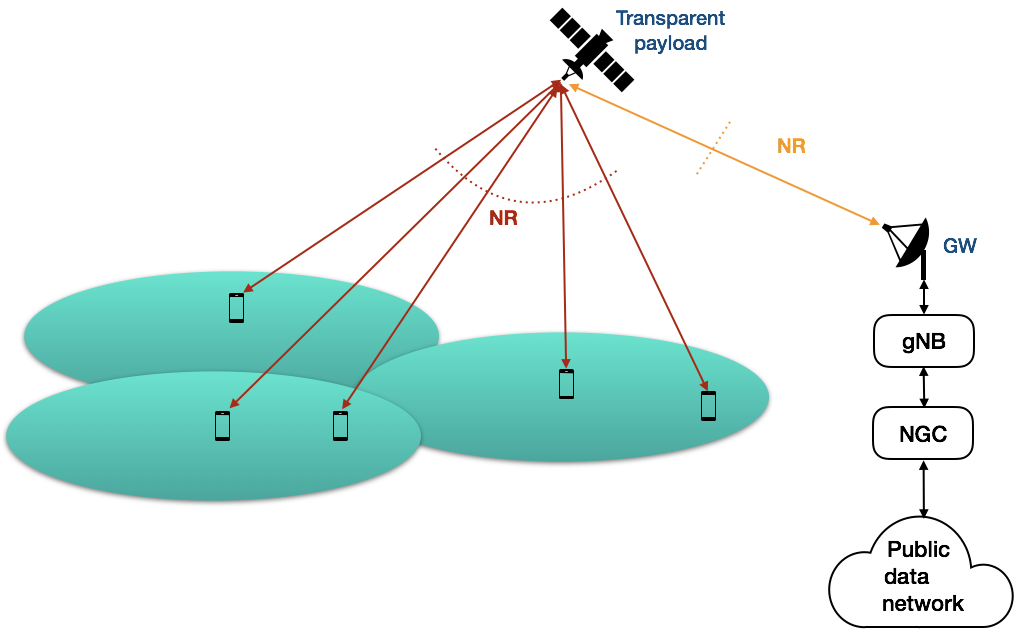}
     }
     \hfill
     \subfloat[A2.\label{fig:ArchitectureA2}]{%
       \includegraphics[width=0.5\textwidth]{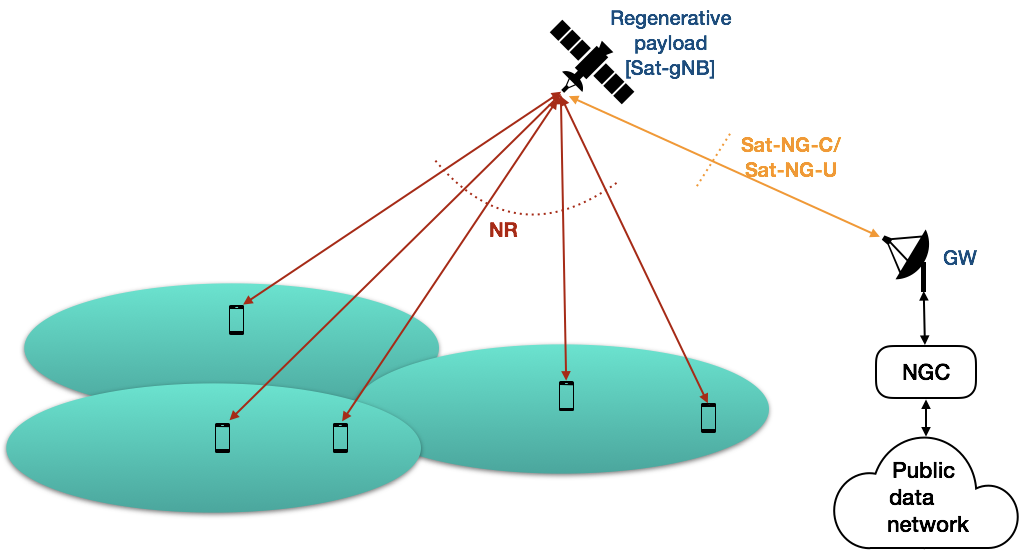}
     }
     \caption{Architecture options with direct user access link.}
     \label{fig:ArchitectureDirect}
\end{figure}

\begin{figure}[t!]
     \subfloat[A3.\label{fig:ArchitectureA3}]{%
       \includegraphics[width=0.5\textwidth]{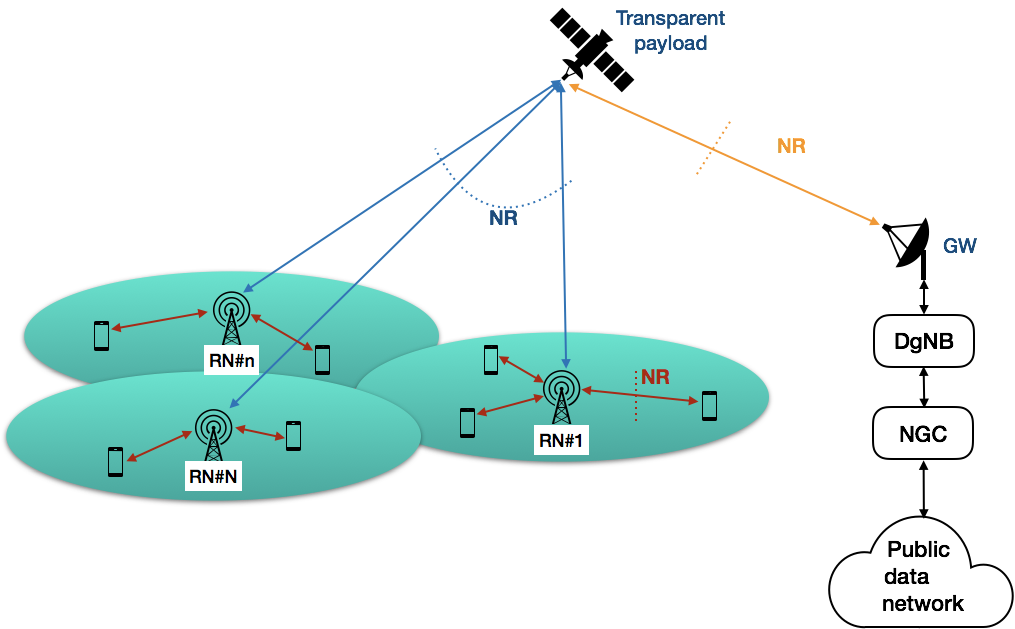}
     }
     \hfill
     \subfloat[A4.\label{fig:ArchitectureA4}]{%
       \includegraphics[width=0.5\textwidth]{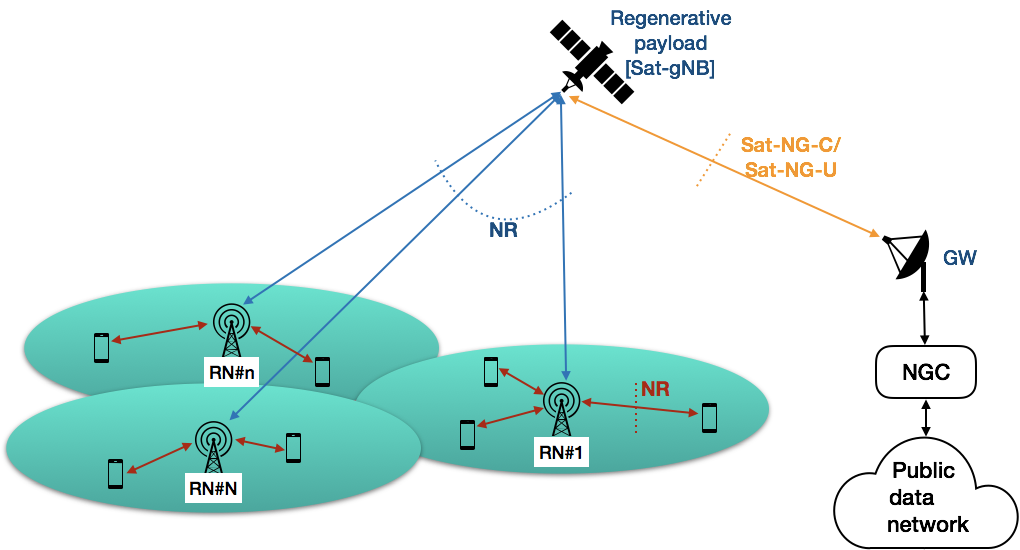}
     }
     \caption{Architecture options with Relay Nodes providing the user access link.}
     \label{fig:ArchitectureRN}
\end{figure}

Non-Terrestrial Networks and, thus, SatCom systems can bring significant benefits to future 5G services thanks to both their wide area service coverage and the significantly reduced vulnerability to physical attacks or natural disasters. In terms of system deployment, both stand-alone 5G SatCom and integrated satellite-terrestrial solutions can be envisaged. This aspect is reflected in the architecture options currently being discussed within 3GPP for NTN, which serve as basis for the analysis performed in the following. In particular, these options can be categorised based on either the type of satellite payload, \emph{i.e.}, transparent or regenerative, and the type of user access link, \emph{i.e.}, direct or through an on-ground Relay Node (RN), as shown in Figures~\ref{fig:ArchitectureDirect}~-~\ref{fig:ArchitectureRN}. It shall be noted that, depending on the satellite altitude, there could be one or more satellites providing on-ground 5G services.

With respect to the direct access scenarios (A1-A2) in Figure~\ref{fig:ArchitectureDirect}, the user access link directly involves the satellite(s) and the on-ground mobile User Equipments (UEs) by means of the New Radio (NR) air interface. This air interface, which is described in the next sections, is currently specifically designed for terrestrial systems and, thus, it is of outmost importance to assess the impact of typical satellite channel impairments, \emph{e.g.}, large delays and Doppler shifts (see Section~\ref{sec:SatImpairments}), on both the Physical layer (PHY), \emph{e.g.}, subcarrier spacing in the NR waveform, and PHY/MAC procedures, \emph{e.g.}, Random Access or Timing Advance. As for the feeder link, the air interface to be implemented depends on the type of satellite payload. On the one hand, in case a transparent satellite is implemented as in Architecture A1, the system gNB(s) is conceptually located at the Gateway (GW) providing the connection towards the Next Generation Core network (NGC) and the public data network. In this architecture, the air interface between the satellite(s) and the GW is again provided by the terrestrial NR, for which the impact of the different satellite channel impairments has to be assessed. On the other hand, when we assume a regenerative payload as in Architecture A2, the gNB is implemented on the satellite, while the GW simply provides the connection towards the NGC and public data network. This architecture is clearly more complex and has a higher cost, but it also allows to significantly reduce the propagation delays and, thus, to ease the modifications (if any) that might be needed to the NR PHY and MAC procedures, which can be directly terminated on the on-board gNB instead of requiring to go down to the GW. The link between the gNB and the NGC can use any suitable air interface, \emph{e.g.}, the SoA DVB-S2X air interface, \cite{DVBS2X}, or an adapted version of the New Radio air interface between gNB and NGC, \emph{i.e.}, NG-C or NG-U. In the following this general air interface is referred to as Sat-NG-C/Sat-NG-U, for the control and user plane information, respectively. It is also worthwhile noting that NR systems foresee the implementation of the functional split concept in the gNB. In particular, the gNB lower layers (namely, up to layer 3) can be implemented in a distributed unit (gNB-DU), which in our case would be located on the satellite, while the remaining layers can be implemented in a centralised unit (gNB-CU), which would be conceptually located at the system GW as for Architecture A1. The air interface between the centralised and distributed units of a gNB is the X1 air interface and the current 3GPP specifications highlight that it is an open air interface, as long as specific signalling operations are guaranteed. Thus, the air interface between a gNB-DU on the satellite and its corresponding gNB-CU at the GW might be implemented by means of existing SatCom standards, as, for instance, DVB-RCS(2), \cite{DVBRCS2}, or DVB-S2X, \cite{DVBS2X}.

When considering architectures A3 and A4, the user access link is provided by on-ground Relay Nodes, to which the backhaul connection is guaranteed by the satellite(s), as shown in Figure~\ref{fig:ArchitectureRN}. NR Relay Nodes, and the related air interfaces, are still under definition within 3GPP. However, we can fairly assume that they will have a similar behaviour to that of LTE RNs. Based on this assumption, the following architectural and operational aspects shall be highlighted: i) the RN is connected to a Donor gNB (DgNB), which provides the connection towards the NGC; ii) the RN can terminate procedures and air interfaces up to Layer 3; iii) the air interface on the user access link (RN-UE) is a normal NR air interface; and iv) the air interface on the backhaul link (RN-DgNB) is a modified version of the NR air interface, for which, however, the only differences are in some Radio Frequency characteristics and minimum performance requirements. Based on these assumptions, the RN basically acts as a traditional UE from the DgNB point of view, while it is seen as a traditional gNB from the UEs within its coverage. In addition to this, we can also state that the air interface on both the user access (RN-UEs) and backhaul (RN-DgNB) links is the NR air interface described in \cite{3GPP_38211}. This architecture is clearly more complex with respect to the direct access scenarios due to the introduction of a certain number $N$ of RNs, which shall be managed by $M$ DgNBs with $N\geq M$, which also increases the overall system cost. However, since the RNs are equivalent to gNBs from the users' perspective, the user access link is now a traditional terrestrial link for which no modifications are needed. The impact of typical satellite channel impairments thus has to be assessed on the backhaul link only, since it is implemented by means of the NR air interface as previously highlighted. With respect to the feeder link between the satellite(s) and the system GW, as for the direct access scenarios, we can have either a NR or a NG-C/NG-U air interface, depending on whether the satellite payload is transparent or regenerative, respectively, and the same observations, \emph{e.g.}, functional split, as provided for options A1-A2 apply.

In the following, we consider architectures A1 and A3, \emph{i.e.}, direct access or access through RNs with transparent payload satellites, in two scenarios for enhanced Mobile Broadband (eMBB) and NarrowBand-Internet of Things (NB-IoT) services, respectively, and discuss the main technical challenges that have to be coped with for the NR air interface due to typical satellite channel impairments. For both scenarios, we assume a Frequency Division Duplexing (FDD) framing structure.

\section{Scenarios}
\subsection{enhanced Mobile BroadBand (eMBB)}
In this scenario, the system aims at providing broadband connectivity to mobile users by means of architecture option A3, \emph{i.e.}, a transparent payload satellite providing backhaul connectivity to on-ground RNs that realise on-ground NR cells. In particular, based on the previous discussion, the following assumptions hold for this scenario if not otherwise specified: i) the air interface on the user access (RN-UE) and backhaul (RN-gNB) links is NR; ii) the RN operates as a normal gNB from the UEs' perspective and, thus, no modifications are needed on the user access link; and iii) the RNs are in fixed positions. With respect to the satellite(s) deployment, we assume a Geostationary Earth Orbit (GEO) with altitude $h_{sat}=35786$ km, system operating in Ka- or S-band. In particular, for Ka-band systems we assume an operating frequency in the range $19.7-21.2$ GHz for the downlink and $27.5 - 30.0$ GHz for the uplink, while in S-band we assume to be working at $2170-2200$ MHz in the downlink and $1980-2010$ MHz in the uplink. It shall be noted that these frequency ranges are proposed in \cite{3GPP_38811} as deployment options for the evaluation of NR-based SatCom systems. The deployment in other frequency bands is clearly possible as long as the national, regional, and/or international spectrum requirements from the radio regulations are met.

\subsection{NarrowBand-IoT (NB-IoT)}
In the second scenario, we focus on massive Machine Type Communications (mMTC) and, in particular, on NB-IoT. To this aim, we assume a direct access architecture with transparent satellite payload, \emph{i.e.}, architecture option A1. Since IoT applications are extremely sensitive to propagation delays and the IoT terminals have stringent battery life requirements, a GEO satellite system is not a feasible option due to the extremely large delays and path loss values. Thus, in the following, we focus on non-GEO systems and, in particular, on a mega-constellation of Low Earth Orbit (LEO) satellites that have an altitude between $h_{sat}=600$ and $h_{sat}=1500$ km. With respect to the operational frequency, we consider the S-band deployment, \emph{i.e.}, $2170-2200$ MHz downlink and $1980-2010$ MHz uplink. Differently from the eMBB case, in this scenario there is no RN providing the user access link and, thus, the impact of the satellite channel impairments described in the next section can be extremely deep on the traditionally terrestrial NR air interface. 

\section{Satellite channel impairments}
\label{sec:SatImpairments}
In order to assess the feasibility of NR techniques and procedures in the eMBB and NB-IoT scenarios previously outlined, in this section we discuss the typical satellite channel impairments that might have an impact on the NR PHY and MAC layers, as large Doppler shifts and propagation delays. 

\subsection{Delay}
With respect to the propagation delay, we have to consider both the one-way propagation delay and the Round Trip Time (RTT), depending on the type of procedure we are considering, \emph{i.e.}, whether the whole procedure or a specific step can be terminated at the gNB or it requires an interaction with the NGC. In the following, the RTT is approximated by twice the propagation delay between the transmitter and the receiver, since the signal processing time in a Satellite Communication context can be assumed negligible with respect to the propagation delay. In order to estimate the propagation delay in the considered scenarios, we further assume to be in a pessimistic scenario in which the transmitter and the receiver are not perfectly aligned and, thus, they have different elevation angles. The overall RTT can thus be computed as follows:
\begin{equation}
\label{eq:RTT}
    RTT \approx 2T_{owp} = 2\frac{d_{GW-Sat}\left(\vartheta_{GW}\right) + d_{Sat-RX}\left(\vartheta_{RX}\right)}{c}
\end{equation}
where $T_{owp}$ is the one-way propagation delay, $d_{GW-Sat}$ the distance between the GW and the satellite as a function of its elevation angle $\vartheta_{GW}$, $d_{Sat-RX}$ the distance between the satellite and the receiver (\emph{i.e.}, the RN or the UE for the eMBB or NB-IoT scenario, respectively) as a function of its elevation angle $\vartheta_{RX}$, and $c$ the speed of light. In the following, the system GW is assumed to be at $\vartheta_{GW}=5^{\circ}$ elevation angle, while the minimum elevation angle for both the UEs and RNs is assumed to be $\vartheta_{RX}=10^{\circ}$. The single paths distances and the related delays between the satellite and both the UE/RN and the GW are listed in Table~\ref{tab:Delays}, where we considered both the minimum and maximum satellite altitude in the NB-IoT scenario. Based on these values, the one-way propagation delay, \emph{i.e.}, RN (UE) to gNB (DgNB), and RTT, \emph{i.e.}, RN (UE) to gNB (DgNB) and back to the RN (UE), shown in Table~\ref{tab:Delays2} can be obtained. It can be noticed that, as expected, for the eMBB scenario the propagation delay might be an issue for all procedures and steps, since it is several orders of magnitude above typical terrestrial networks delays. With respect to the NB-IoT scenario, the impact of the delays shall be evaluated on a case-by-case basis also taking into account the type of communication that the considered procedure is requesting, \emph{i.e.}, towards the gNB or the NGC.

\begin{table}[t!]
\renewcommand{\arraystretch}{1.3}
\caption{Worst-case single path distances and related delays for the considered system architecture.}
\label{tab:Delays}
\centering
\begin{tabular}{|c|c|c|c|}
\hline
\multicolumn{4}{|c|}{ \bfseries eMBB scenario: GEO $h_{sat}=35786$ km}\\
\hline
\bfseries Elevation angle & \bfseries Path & \bfseries Distance [km] & \bfseries Delay [ms] \\
\hline
RN: $\vartheta_{RN}=10^{\circ}$ & Sat-RN & $40586.07$ & $\approx 135.28$\\
\hline
GW: $\vartheta_{GW}=5^{\circ}$ & Sat-GW & $41126.72$ & $\approx 137.09$\\
\hline
\hline
\multicolumn{4}{|c|}{ \bfseries NB-IoT scenario: LEO $h_{sat}=600$ km}\\
\hline
\bfseries Elevation angle & \bfseries Path & \bfseries Distance [km] & \bfseries Delay [ms] \\
\hline
UE: $\vartheta_{UE}=10^{\circ}$ & Sat-UE & $1932.25$ & $\approx 6.44$\\
\hline
GW: $\vartheta_{GW}=5^{\circ}$ & Sat-GW & $2329.03$ & $\approx 7.76$\\
\hline
\multicolumn{4}{|c|}{ \bfseries NB-IoT scenario: LEO $h_{sat}=1500$ km}\\
\hline
\bfseries Elevation angle & \bfseries Path & \bfseries Distance [km] & \bfseries Delay [ms] \\
\hline
UE: $\vartheta_{UE}=10^{\circ}$ & Sat-UE & $3647.55$ & $\approx 12.16$\\
\hline
GW: $\vartheta_{GW}=5^{\circ}$ & Sat-GW & $4101.72$ & $\approx 13.67$\\
\hline
\end{tabular}
\end{table}

\begin{table}[t!]
\renewcommand{\arraystretch}{1.3}
\caption{One-way propagation delay and RTT for the eMBB and NB-IoT scenarios.}
\label{tab:Delays2}
\centering
\begin{tabular}{|c|c|c|}
\hline
\bfseries Scenario & \bfseries One-way [ms] & \bfseries RTT [ms] \\
\hline
\hline
eMBB & $\approx 272.37$ & $\approx 544.75$\\
\hline
NB-IoT $h_{sat}=600$ km & $\approx 14.2$ & $\approx 28.4$\\
\hline
NB-IoT $h_{sat}=1500$ km & $\approx 25.83$ & $\approx 51.66$\\
\hline
\end{tabular}
\end{table}

\subsection{Doppler shift}
The Doppler shift consists in the change in the carrier frequency due to the relative motion between the satellite and the user terminal. The maximum target user mobility in NR requirements is set to $500$ km/h for frequencies below $6$ GHz and it is defined as the maximum NR speed with respect to the serving gNB at which the NR can be served with a specific guaranteed Quality of Service (QoS). Assuming a carrier frequency $f_c = 4$ GHz, and by reminding that the Doppler shift can be computed as $f_d = (v\cdot f_c)/c$, where $v$ is the relative speed between the transmitter and the receiver, it is possible to obtain a maximum Doppler shift $f_d = 1.9$ kHz. If we focus on the considered SatCom scenarios in Ka-band, assuming a carrier frequency $f_c=20$ GHz the maximum Doppler shift for a $500$ km/h relative user mobility becomes approximately $9.3$ kHz, which is clearly significantly above the maximum values assumed for terrestrial NR systems. 

When considering satellite communications, the Doppler shift can be caused by the satellite movement on its orbit and the user terminals' mobility on ground. It shall be noticed that, in the considered eMBB scenario, we are considering GEO systems serving fixed on-ground Relay Nodes and, thus, the Doppler shift can be assumed to be negligible. On the other hand, when we consider the NB-IoT scenario, we have non-GEO satellites serving moving user terminals and, thus, the Doppler shift can introduce significant frequency shifts with respect to those expected in terrestrial NR systems. This could deeply impact the frequency synchronisation of the resources used to transmit through the NR air interface, as discussed in the following sections. In order to assess the impact of Doppler shifts on the NR specifications, we refer to \cite{Intro_UniBo}, where the authors provided a closed-form expression for the Doppler shift as a function of the satellite orbital velocity (relative to the user terminal) and the elevation angle:
\begin{equation}
\label{eq:Doppler}
    f_d(t) = \frac{f_c\cdot \omega_{sat}\cdot R_E\cdot \cos\left(\vartheta_{UE}(t)\right)}{c}
\end{equation}
where $\omega_{sat}=\sqrt{GM_E/{\left(R_E+h_{sat}\right)}^3}$ is the satellite orbital velocity, $R_E$ the Earth radius, $G = 6.67\cdot 10^{-11}$ Nm$^2$/kg$^2$ the Gravitational constant, and $M_E = 5.98\cdot 10^{24}$ kg the Earth mass.

\section{Technical Challenges Analysis}

\subsection{eMBB scenario}
\label{sec:eMBB}
In this section, the physical layer for NR systems is described and the impact of the previously outlined satellite channel impairments is assessed for the eMBB scenario.
\subsubsection{PHY}
In general, the NR waveform is OFDM-based for eMBB scenarios. The baseline waveform is CP-OFDM, which is similar to LTE, is designed to have improved spectral utilization by means of an extended numerology flexibility and adaptability, among the others, \cite{3GPP_38211}. The DFT-s-OFDM based waveform is also supported in the uplink, complementary to CP-OFDM waveform at least for eMBB for up to $40$ GHz bandwidths. In general, the specifications state that CP-OFDM waveforms can be used for a single-stream and multi-stream (\emph{i.e.}, MIMO) transmissions, while DFT-s-OFDM waveform is limited to a single stream transmissions targeting link budget-limited cases. In terms of framing and numerology, NR allows for a significantly improved flexibility with respect to legacy LTE systems. In particular, the frame structure is designed to provide flexibility in the choice of subcarrier spacing, FFT size, subframe duration, and CP length. The subframe duration for a reference numerology with subcarrier spacing $(15\cdot 2^n)$ kHz is exactly $(1/2^n)$ ms, with $n=0,1,\ldots,5$, \emph{i.e.}, from $15$ kHz to up to $480$ kHz subcarrier spacings. The working assumption is that symbol-level alignment across different subcarrier spacings with the same CP overhead is assumed within a subframe duration in a NR carrier. The Physical Resource Block (PRB) is defined as $12$ subcarriers.

With respect to the modulation schemes, QPSK, 16QAM, 64QAM, and 256QAM (with the same constellation mappings as in LTE) are supported for both the downlink and the uplink. $\frac{\pi}{2}$-BPSK in also supported in the uplink for DFT-s-OFDM waveforms. The channel coding scheme is different depending on the type of information to be transmitted, \emph{i.e.}, whether it is Control Plane (CP) or User Plane (UP). For the UP in eMBB scenarios, flexible LDPC codes are implemented as a single channel coding scheme for medium to long block sizes, while in the CP, in which short codewords have to be transmitted, the standardised channel coding scheme is Polar Coding (except for very small block lengths where repetition/block coding may be preferred). One of the critical aspect in this framework is related to the employed modulation and coding formats. First of all, due to the low Signal-to-Noise Ratio (SNR) levels in SatCom, it might be necessary to extend the modulation and coding schemes to meet the reliability requirements of terrestrial NR systems. In addition to this, in typical terrestrial cellular systems, the gNB selects the most appropriate modulation and coding scheme based on the channel quality indicator reported by the UE. Due to the large delay of satellite systems, channel information provided by the UE could be not updated, leading to a suboptimal use of the channel with consequently lower spectral efficiency. In this framework, it is thus necessary to: i) improve the channel information; and. ii) optimise the adaptive modulation and coding techniques for the UE, the RN, and the gNB packets transmissions. As a possible solution, the RN could have an active role, optimising both the user access and the backhaul air interfaces, which would require a new approach to exchange channel information between these nodes.

\subsubsection{Candidate Waveforms and Performance Evaluation}
Candidate waveforms that have been studied for NR, apart from the selected CP-OFDM and DFT-s-OFDM schemes, include Filtered OFDM (f-OFDM), Windowed OFDM (W-OFDM), Filter Bank Multicarrier (FBMC), Universal Filtered Multicarrier (UFMC), and Generalised Frequency Division Multiplexing (GFDM). An analysis in \cite{BookChapter5GWaveforms} for the terrestrial scenario shows that, while these waveforms achieve lower Out-Of-Band-Emissions (OOBE), they mostly suffer from higher PAPR and higher computational complexity. However, they provide other advantages such as higher time-frequency efficiency, support for short-burst traffic, fragmented spectrum, and greater tolerance to timing and frequency offsets. The work done in \cite{ArtFilteredOFDM} shows that f-OFDM is a flexible waveform that appears to be the most promising candidate waveform for NR cellular networks.

\begin{figure}[t!]
     \subfloat[ \label{fig:Figure1}]{%
       \includegraphics[width=0.5\textwidth]{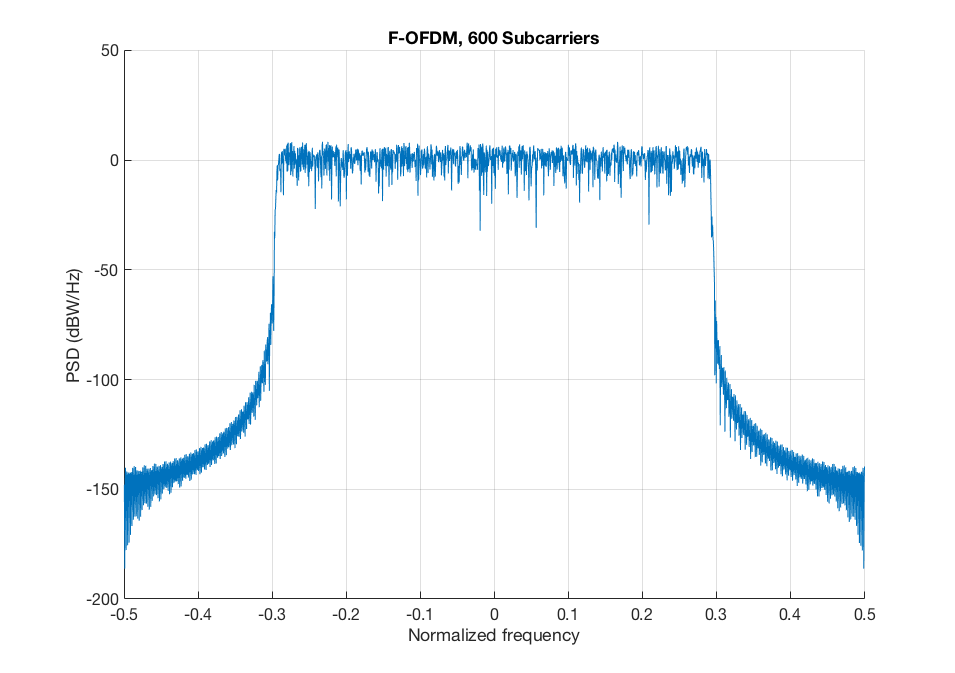}
     }
     \hfill
     \subfloat[ \label{fig:Figure2}]{%
       \includegraphics[width=0.5\textwidth]{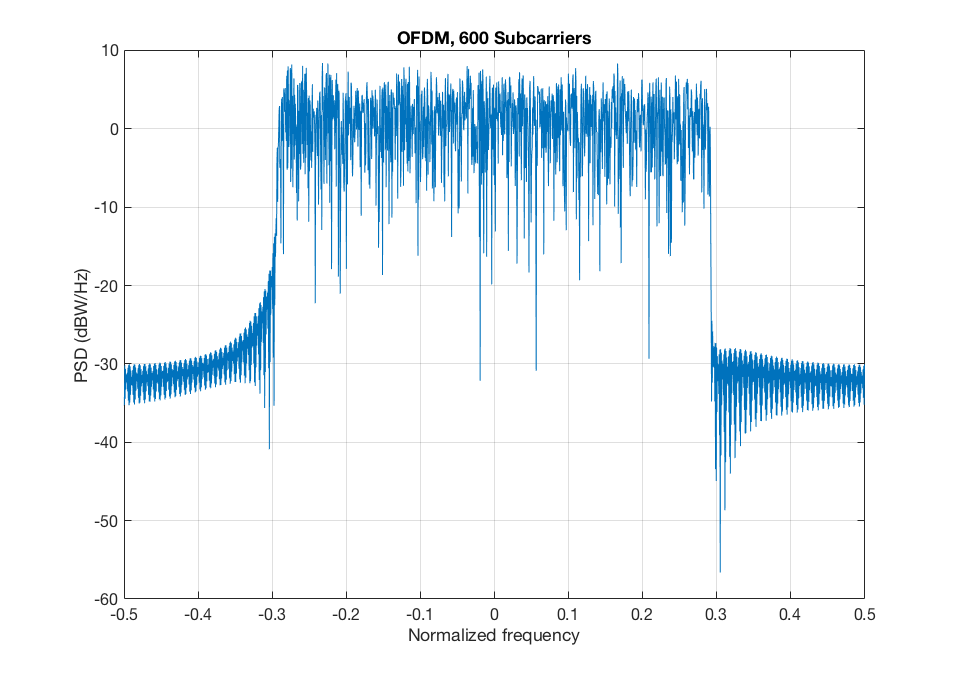}
     }
     
  \subfloat[ \label{fig:Figure3}]{%
       \includegraphics[width=0.5\textwidth]{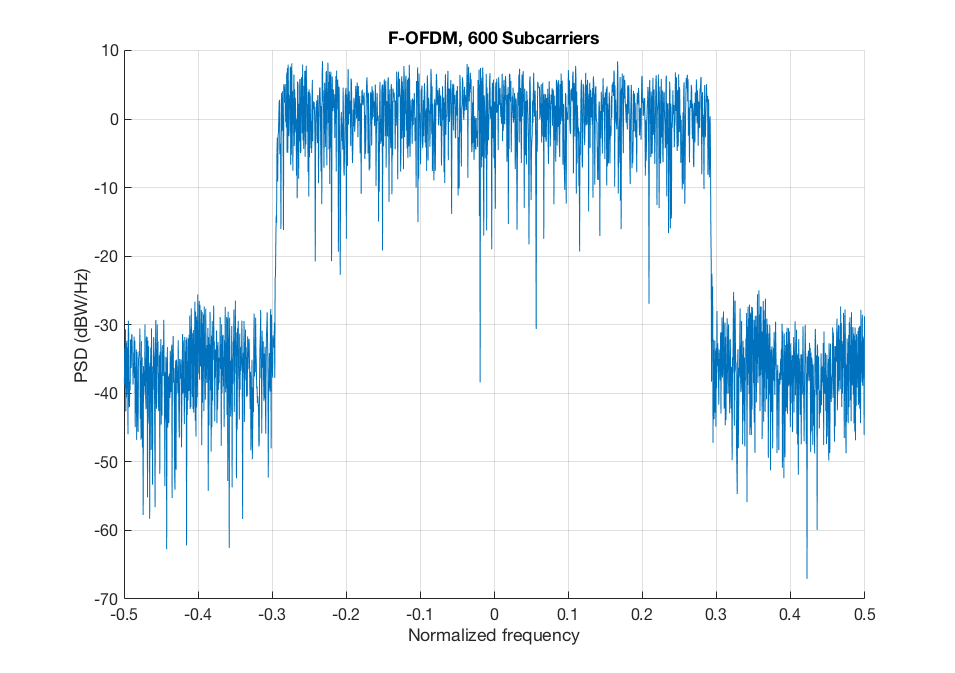}
     }
     \hfill
     \subfloat[ \label{fig:Figure4}]{%
       \includegraphics[width=0.5\textwidth]{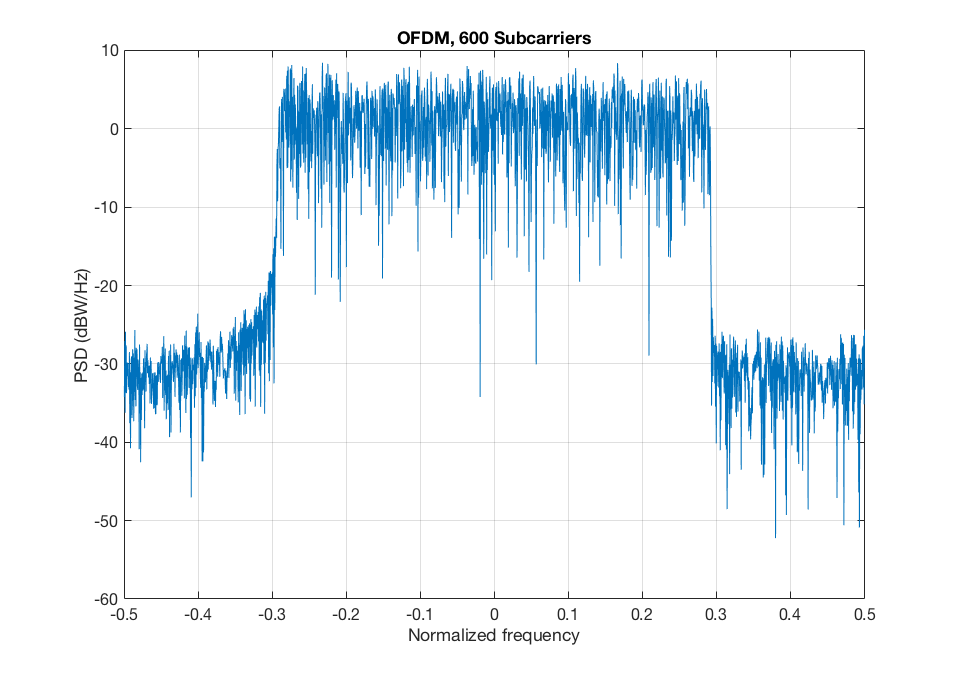}
     }
     \caption{f-OFDM preliminary performance analysis compared to legacy OFDM.}
     \label{fig:WaveformPerformance}
\end{figure}

\begin{figure}[t!]
        \centering
        \includegraphics[width=0.6\textwidth]{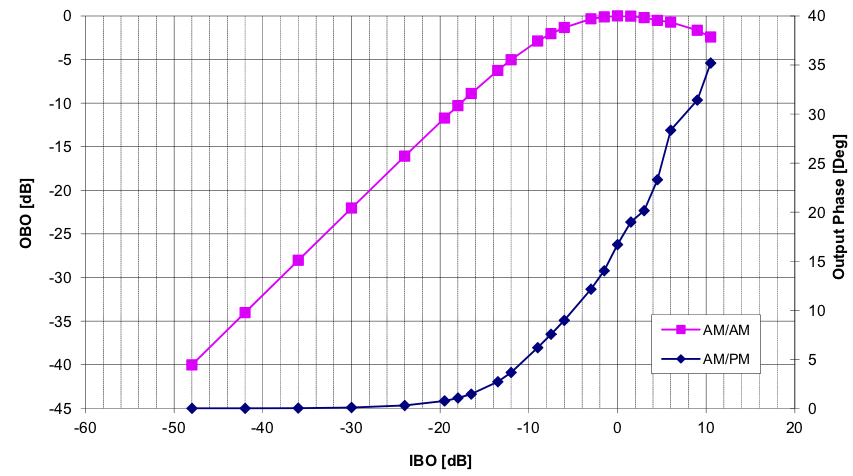}
        \caption{TWTA response for Ka-band.}
        \label{fig:Figure5}
\end{figure}

As discussed before, the OFDM-based NR waveform allows for scalable subcarrier spacing, filtering, subframe duration, CP length and windowing. Given that the mobility of the satellite with regards to the RN is not a major issue for a GEO-based deployment scenario, we focus on optimising the satellite waveform design with respect to TWTA nonlinearities. Figures~\ref{fig:Figure1}-\ref{fig:Figure4} illustrate a preliminary analysis of the performance of F-OFDM in comparison to the legacy OFDM in a satellite nonlinear channel. We use a basic OFDM design for this analysis, having an FFT size $N=1024$ with $600$ used subcarriers (loaded with 64QAM data symbols) and a CP length $72$, in an AWGN channel. A filter with a rectangular frequency response (\emph{i.e.}, a sinc impulse response) is used with a filter length $L=513$ on the OFDM signal to produce the F-OFDM signal. The low-pass filter is realized using a window which effectively truncates the impulse response and offers smooth transitions to zero on both ends. The satellite nonlinear channel is modelled based on the conventional TWTA response for Ka-band specified in \cite{SXChannel} and shown in Figure~\ref{fig:Figure5}. Figure 1 shows the significant improvement in OOBE that is achievable by using F-OFDM in a linear channel, wherein the OOBE suppression is about $150$ dB in contrast to the $30$ dB achieved by OFDM as shown in Figure~\ref{fig:Figure2}. However, this improvement is degraded in a satellite nonlinear channel, even for an input back-off, $IBO=20$ dB, with F-OFDM degrading back to the legacy OFDM performance as shown in Figures~\ref{fig:Figure3}-\ref{fig:Figure4}. This emphasizes the high sensitivity of F-OFDM to non-linear distortion. This is because of the high level filtering applied, which also results in higher PAPR.  
Future work will incorporate nonlinear compensation techniques in studying how to improve and optimise the performance of F-OFDM over the satellite nonlinear channel.

\subsubsection{PHY/MAC Procedures}
In this section, we review the most critical PHY and MAC layer procedures of NR systems for the eMBB scenario. In particular, the impact of the large delays and Doppler shifts previously introduced on these procedures is assessed and potential solutions are proposed. To this aim, as already mentioned in the previous sections, it shall be noted that in the eMBB scenario we are focusing the backhaul link between the RN and the DgNB, which are fixed and communicate through a GEO satellite. As a consequence, the Doppler shift does not introduce any issue, while the very large propagation delays might have a strong impact.

\subsubsection*{Timing Advance}
The Timing Advance (TA) is a time offset used in the timing adjustment procedure that is performed by the UEs with support form the gNB, \cite{3GPP_38211,3GPP_38213}. In particular, the TA is a negative offset that informs the UE on the correct uplink transmission timing so as to guarantee that all of the uplink frames received by the gNB from its users are aligned with the corresponding downlink frames. Since the time-frequency resource blocks are assigned to the UEs on a $0.5$ ms slot, this procedure ensures that resources assigned for user data in a time slot are not going to interfere with control or estimation signals that can be in the same time-frequency locations in the following time slots. The TA value, $T_{TA}$, is estimated by the gNB during the Random Access procedure when the NR UE is trying to establish an RRC connection, and then the value is provided to the UE within the response to the access request. After the initial timing adjustment during the RRC connection establishment, all the following adjustments are provided by the gNB as a difference with respect to the current TA value. The UE shall adjust the timing of its uplink transmission at sub-frame $n+6$ for a timing advancement command received in sub-frame $n$ and, in addition, the UE also has a reconfigurable timer, the \emph{timeAlignmentTimer}, which is used to control for how long the UE is considered time aligned in the uplink and can be set up to $10.24$ s, \cite{3GPP_38321,3GPP_38331}.

From NR specifications, the TA value is obtained as $T_{TA}=T_C \left(N_{TA} + N_{TA,offset}\right)$, where $N_{TA,offset}$ is a parameter depending on the considered frequency band and $N_{TA}=T_A\cdot 16\cdot 64\cdot/{2^{\mu}}$, where $T_A=0,1,\ldots,1282$ is the Timing Advance Command provided by the gNB in the response to the Random Access request, $T_C = 1/\left(\Delta f_{max}N_f\right)$ with $N_f=4096$ and $\Delta f_{max}=480$ kHz is the basic NR time unit, and $\mu$ denotes the subcarrier spacing as $2^{\mu}\cdot 15$ kHz. When an FDD framing structure is considered, $N_{TA,offset}=0$ and we can write the overall TA adjustment as $T_{TA}=N_{TA}T_C = T_AT_{TA}^{(single)}$, where $T_{TA}^{(single)}$ is the TA basic unit, \emph{i.e.}, the timing advance value for $T_A=1$. This value depends on the subcarrier spacing and, in particular: i) for $\mu=0$ we have the same subcarrier spacing as in LTE, which leads to $T_{TA}^{(single)}=0.52$ ns and to a maximum TA value $T_{TA}=0.6667$ ms; and ii) for $\mu=5$, $T_{TA} = 0.16\cdot 10^{-2}$ ns and, thus, the maximum TA value is $T_{TA} = 0.0209$ ms. These timing adjustment values correspond to a step in the estimation of the distance between the UE and the gNB equal to $c\cdot T_{TA}/2$, where the division by $2$ is to account for the RTT, which leads to $78.125$ meters for $\mu=0$ and $2.441$ meters for $\mu=5$.

Based on the above description, there are two aspects to be assessed in the SatCom context: i) the maximum allowed TA shall be such that the maximum differential delay between the considered UEs is below its value. In fact, in case this condition is not satisfied, there will be some UEs that will receive a TA command below the value that would actually guarantee the uplink and downlink frame synchronisation at the gNB, which would lead to disruptive interference; and ii) the reconfigurable timer that informs on the validity of the current TA value shall be above the maximum RTT, so as to guarantee that the UEs are considering a valid TA adjustment. With respect to the former timer, it shall be noted that we are considering the eMBB scenario, in which we have a fixed GEO satellite and several fixed RNs on-ground. In general, the maximum allowed TA value should be such that the maximum distance between any couple of RNs does not introduce a differential time delay above it. As previously computed, the maximum allowed TA in the worst-case scenario, \emph{i.e.}, maximum subcarrier spacing, is $T_{TA} = 0.0209$ ms, which leads to a maximum distance of $3.135$ km, while in the best case $T_{TA} = 0.6667$ ms leads to $100$ km. Both these distances can be significantly below the distance between a couple of RN, since we are addressing a GEO system with a single satellite covering a very large portion of the Earth. However, as already highlighted, we are in a fixed scenario both from the transmitter (gNB at the GW) and the receiver (RNs) point of view and, thus, the TA value can be easily set to the specific delay value between each RN and the system GW when the RN is deployed in the coverage area, \emph{i.e.}, through an \emph{ad hoc} system deployment. The TA adjustment can then still be implemented only to take into account the differential adjustments that might be required in order to compensate for the orbit adjustment of the GEO satellites, which in any case will be below the maximum distance of . As for the reconfigurable parameter that informs the RNs on the validity, in the time domain, of the current TA value, also based on the ad hoc system deployment that can be realised, the only aspect to be considered is that this value ensures that the TA adjustment related to the satellite orbit compensation is valid for a sufficient period. Since the maximum value of \emph{timeAlignmentTimer} is equal to $10.24$ s, which is significantly below the one-way and RTT delays outlined in the previous section, we can conclude that there is no specific issue for the NR TA in the considered eMBB scenario.

\subsubsection*{Random Access}
\begin{figure}[t!]
     \subfloat[Contention-based RA.\label{fig:RA_contention-based}]{%
       \includegraphics[width=0.35\textwidth]{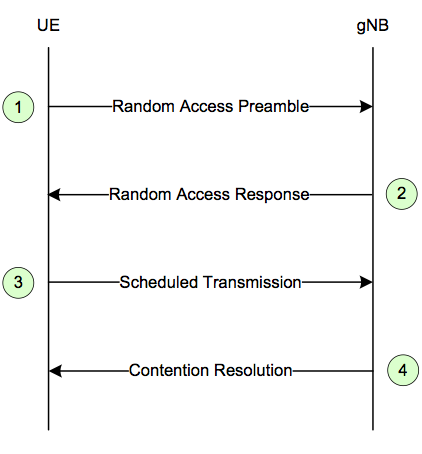}
     }
     \hfill
     \subfloat[Contention-free RA.\label{fig:RA_contention-free}]{%
       \includegraphics[width=0.35\textwidth]{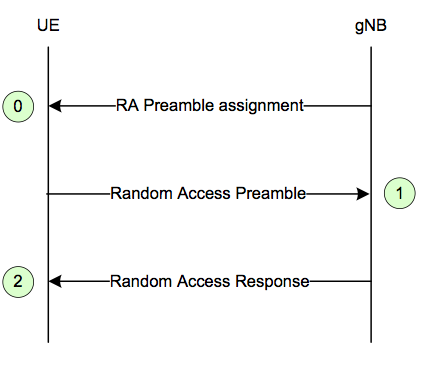}
     }
     \caption{Contention-based and contention-free Random Access procedures in NR.}
     \label{fig:RAprocedure}
\end{figure}

The Random Access (RA) procedure between a RN and its DgNB is the same as that between a NR UE and the gNB and, in particular, it is the same to that implemented in LTE, from an algorithmic point of view, \cite{3GPP_38321}. The RA procedure can be: i) contention-based, when the UE is not yet synchronised or lost its synchronisation; or ii) contention-free, in case the UE was previously synchronised to another gNB. Both procedures rely on the transmission of a random access preamble from the UE to the gNB, which shall be performed on specific time-frequency resources that are indicated by the gNB on the control channels. The two procedures are shown in Figure~\ref{fig:RAprocedure}. Focusing on the contention-based procedure, since the contention-free is included in it involving only the first two steps, the following operations are performed:i) in step 1, the NR UE randomly choses a preamble from a predefined set, also based on preliminary information on the expected amount of resources to be used in the subsequent (if any) step 3, and sends it to the gNB along with a temporary network identifier, which is computed based on the RA preamble as well; ii) in step 2, the gNB responds to the request with a RA Response (RAR) message, which shall be received by the UE within a RA time window between starting after the transmission of the last preamble symbol. The value of this time window is still under discussion within 3GPP standardisation, but we can fairly assume that it will be similar to the LTE time window, which can be set up to $15$ ms. If this counter expires, the UE can attempt a new RA procedure, up to 200 tentatives; and iii) in steps 3 and 4 the NR UE is assigned a final network identifier, subject to the successful resolution of any possible contention. In particular, HARQ is implemented in step 3, with a contention timer up to $64$ subframes, \emph{i.e.}, $64$ ms, and up to $16$ tentatives, \cite{3GPP_38211,3GPP_38213,3GPP_38321,3GPP_38331}. If the UE receives a correct response in Step 4, the procedure is successful and it is now logged to the NR network. It is worthwhile highlighting that the contention-free procedure involves Step 1 and 2 only.

In the considered eMBB scenario, and focusing on the contention-free RA, the UEs in each on-ground cell perform the RA procedure only involving the corresponding RN, which terminates all protocols up to Layer 3, and thus this does not require any modification since the satellite channel does not come into play. As for the contention-based RA, in step 3 and 4 the RN shall contact the NGC, through the DeNB, so as to obtain a final network identifier for the NR UE. In this moment, the delay on the satellite channel is involved and shall be carefully taken into account. However, as previously reported, the contention timer in this phase of the RA procedure is set to up to $15$ ms, which is significantly lower than the RTT between the RN and the NGC for the eMBB scenario. Thus, from the NR UE perspective, the RA procedure might require the extension of the contention resolution timer so as to cope with the RTT for GEO systems. In addition to this, the RA procedure shall also be performed at the beginning of the RN start-up procedure, since, as highlighted in the previous sections, the RN is initially seen by the DgNB as a normal NR UE. In this case, both the RAR window and the contention resolution timer shall be compared with the satellite channel RTT, which is significantly larger than the maximum values foreseen in the NR specifications. Thus, as for the contention resolution timer for the UE RA procedure, in the RN start-up procedure we might need to extend both the contention resolution timer and the RAR time window. However, similarly to the Timing advance procedure, it shall be noted that the RN start-up is only required when the RN is switched-on and connected to its DgNB at the system GW. Thus, the RN start-up procedure can be easily avoided by means of an \emph{ad hoc} RN initialisation and, since both the RN and the GEO satellite are fixed, no further actions will be required in this context. 

\subsubsection*{HARQ}
The NR air interface supports a one-bit per Transport Block (TB), \emph{i.e.}, MAC layer information block, for HARQ-ACK reporting. Both for the uplink and the downlink, multiple parallel Stop-And-Wait (SAW) HARQ processes can be run within each MAC entity in both the gNB and the NR UR, \cite{3GPP_38213,3GPP_38321,3GPP_38331}. The maximum number of processes is $16$ for the downlink HARQ and either $8$ or $16$ for the uplink HARQ, but is shall be noted that these values still have to be confirmed within 3GPP. Each process is asynchronous, \emph{i.e.}, retransmissions can happen at any time relative to the first attempt, thus requiring an HARQ identifier, and adaptive, \emph{i.e.}, transmission attributes are adaptively reconfigured to the channel conditions, which is then exploited by either Incremental Redundancy (IR) or Chase Combining (CC) TB combination approaches. In this context, the critical aspect to be highlighted is that the HARQ procedure is asynchronous in terms of the time slot in which a UE or the gNB can send again a specific TB, which required the association to a specific HARQ identifier, but the time slots in which a UE (gNB) is allowed to send its ACK is tightly specified in the standard. In particular, a UE (RN) receiving a TB with the last symbol in subframe $n-k$ shall send the ACK/NACK information in subframe $n$, where $k$ is an offset in number of subframes indicated from higher layers in the control information and it takes into account the propagation delay between the UE and RN. If this parameter is not specified by the upper layers, by default a UE (RN) that receives the last TB symbol in subframe $n-k$ shall provide the related ACK/NACK in time subframe $n-k+4$. The allowed values for the parameter $k$ are still under discussion and, thus, for the following analysis we focus on the default solution in which the delay between the reception of a TB and the transmission of the related ACK/NACK is $4$ subframe. It shall be noted that this solution corresponds to the legacy LTE HARQ processing time, which takes into account $3$ ms of processing delay for the TB (ACK/NACK) and a $1$ ms propagation delay (for the maximum cell size of $100$ km) for an overall $4$ ms.

The most critical aspect to be considered when implementing NR HARQ procedures over a satellite channel is that the minimum number of parallel processes is actually obtained as a function of the overall HARQ processing time as $N_{HARQ}^{(min)} = T_{HARQ}/TTI$, where $T_{HARQ}$ is the HARQ processing time and $TTI$ ms is the transmission time interval for one TB. In the NR air interface it is highlighted that the HARQ processing time at least includes the delay between the downlink data reception timing to the corresponding ACK/NACK transmission timing and the delay between the uplink grant reception timing to the corresponding uplink data transmission timing. Thus, $T_{HARQ}$ includes the time for the UE (RN) to send the ACK/NACK for the TB received in time subframe $n-k$, \emph{i.e.}, $4$ subframes, and the time for the RN (UE) to send again the data if a NACK was received, \emph{i.e.}, again $4$ subframes, which leads to $T_{HARQ}=8$ subframes, \emph{i.e.}, $8$ ms in FDD framing. Differently from legacy LTE systems, the TTI for NR systems depends on the subcarrier spacing. In particular, when a $15$ kHz subcarrier is considered, the TTI is equal to $1$ ms as in LTE, while for a subcarrier spacing of $2^{\mu}\cdot 15$ kHz, with $\mu=1,2,\ldots,5$, the TTI is $1/2^{\mu}$ ms. Thus, the maximum number of parallel processes might go to up to $128$ with a $480$ kHz subcarrier. However, this is still under discussion with 3GPP RAN meetings and, in particular, for the moment being the maximum number of parallel processes has been set to $16$, \emph{i.e.}, an acceptable value is considered the one obtained with $30$ kHz carriers and $0.5$ ms TTI values.

In the considered eMBB scenario, we have a significantly larger propagation delay with respect to the terrestrial case. In particular, for the computation of the maximum number of HARQ processes, while we can still assume a $3$ ms processing delay for the TB (ACK/NACK), the propagation delay is the one provided in Table~\ref{tab:Delays2}. Thus, the maximum number of parallel HARQ processes in the considered system is $N_{HARQ} \approx 2\times 277.37/TTI$ that, with $TTI=1$ ms, leads to $555$ processes. This has a two-fold impact on the system: i) the soft-buffer size of the UE, which is given by $N_{buff}\propto N_{HARQ}TTI$; ii) the bit-width of DCI fields would have to be increased to at least $10$ bits so as to guarantee the possibility to identify $N_{HARQ}$ processes. On one hand, increasing the UE buffer size can be very costly, while on the other hand larger bit-widths of DCI field would lead to large DL control overhead. To cope with the above issues, different solutions can be proposed: i) increasing the buffer size to cope with the large number of HARQ processes; ii) increase the number of HARQ processes, by maintaining the buffer size under control, by using a 2 bit ACK, \cite{3GPP_HARQref}, to inform the transmitter on how close the received packet is to the originally transmitted one. Therefore, the number of retransmission will be reduced, because the transmitter can add the redundant bits according to the feedback information; iii) reducing the number of HARQ processes and the buffer size, which also reduces the system throughput; and iv) not implementing the HARQ protocol, which requires solutions to solve issues related to colliding/non-decodable packets. These solutions all have an impact on either the overall system cost or its throughput and, therefore, shall be carefully assessed.

\subsection{NarrowBand-IoT}
\label{sec:NBIOT}
LTE releases have provided progressively improved support for Low Power Wide Area Network (LPWAN). In fact, Release 13 EC-GSM-IoT\cite{3gpp:45820} and LTE-eMTC\cite{3gpp:45820} aim to enhance existing GSM and LTE networks, for better serving IoT use cases. A third technology, the so called Narrowband IoT (NB-IoT)\cite{tech:qualcomm_narrowband}, shares this objective as well. While it is designed to achieve excellent coexistence with legacy GSM and LTE technologies, NB-IoT is actually a new technology and, as a consequence, not fully backward compatible with existing 3GPP devices\cite{journal:wang_primer}. According to 3GPP, NB-IoT aims at offering\cite{conference:ratasuk_overviewNBIoT}:
\begin{itemize}
        \item Ultra-low complexity devices to support IoT applications.
        \item Improved indoor coverage of $20\,dB$ compared to legacy GPRS, corresponding to a Maximum Coupling Loss (MCL) of $164\,dB$ while supporting a data rate of at least $160\,bps$ for both uplink and downlink.
        \item Support of massive number of low-throughput devices (at least 52547 devices within a cell-site sector).
        \item Improved power efficiency (battery life of 10 years) with battery capacity of 5 Wh and transmission power depending on the terminal power class (\emph{e.g.}, $20$ dBm for Power Class 5 and $23$ dBm for Power Class 3).
        \item Exception report latency of $10\,s$ or less for 99\% of the devices.
\end{itemize}
Possible applications varies from  smart metering, smart cities and buildings, environmental and agriculture monitoring, up to animal/people tracking.

In the following subsections, at first we are going to highlight a list of the most important challenges for this technology to be used in the considered architecture/scenario (Sec. \ref{s:nbiotchallenges}), then, some useful information related to the NB-IoT physical layer waveforms and, in particular, an assessment on the effect of differential Doppler and frequency offset on NB-IoT transmission are described, and finally  an investigation on the timing constraints and on the procedures of NB-IoT are addressed.

\subsubsection{Main challenges in the use of NB-IoT over LEO satellites}
\label{s:nbiotchallenges}
The aim of the current section is to highlight the principal challenges to be faced concerning the integration of NB-IoT over satellite links. In particular, the focus is on LEO satellites due to latency and link budget constraints required by NB-IoT. While the following item list is not completely exhaustive of all the possible issues to be tacked in the whole study, it includes the most important parameters: 
\begin{enumerate}
        \item \textbf{Latency:} even if the latency constraints in NB-IoT are relaxed\cite{conference:ratasuk_overviewNBIoT}, some timers coming from LTE architecture have to be taken into account into the investigation. In particular the study on the 3GPP standards highlighted the presence of timers in the Random Access procedure, shown in Figure~\ref{fig:RAprocedure} and RRC procedures that might be incompatible with SatCom channel Round Trip Time (RTT) delays, which are:
        \begin{itemize}
                \item RRC Timers procedure
                \item RAR time window size
                \item Contention Resolution window size
                \item Timing Advance (TA)
                \item HARQ process
        \end{itemize}
        
        \item \textbf{Doppler Effect and Phase Shift:} due to the NB-IoT frame structure, with really narrow and close subcarriers, the Doppler and phase impairments characteristics of a satellite communication channel, in particular when LEO or MEO satellites are considered, could prevent a successful transmission.
        In particular, the differential Doppler and Carrier Frequency Offset (CFO) amongst users is a potential source of degradation for Scenario A.        
        \item \textbf{Battery Life:} NB-IoT requirements suggest a battery life around 10 years. The longer RTT, characteristic for satellite communications, will imply longer wake up period for devices in order to perform access procedures and data transmission. Furthermore higher power could be needed in order to close the link. These issues could prevent the long duration of batteries.
          \item \textbf{Link Budget:} power constraints of Satellite, eNodeB (eNB) and NarrowBand User Equipment (nUE or UE) must be considered for the feasibility of GEO and LEO feeder and user links, on both forward and return paths.      
\end{enumerate}
While the link budget is an important aspect to be taken into account in order to guarantee a reliable transmission, the following analysis assumes the system to be modelled so that the link can be closed. This is justified, for example, by the use of proper satellite antenna gains.

\subsubsection{NB-IoT Physical Layer}

An operator can deploy NB-IoT inside an LTE carrier by allocating one of the Physical Resource Blocks (PRBs) of $180\,kHz$ to NB-IoT, or in GSM spectrum, reducing the deployment costs. Three different modes of operation have been defined\cite{journal:yu_uplink_2017}:
\begin{itemize}
        \item \textit{In-Band}: NB-IoT is deployed inside an LTE carrier. The NB-IoT carrier consists of one resource block of $180\,kHz$. In this case, LTE and NB-IoT share transmit power at the eNB.
        \item \textit{Guard-Band}: the NB-IoT channel is placed in a guard band of an LTE channel. The NB-IoT downlink can share the same power amplifier as the LTE channel.
        \item \textit{Stand-Alone}: NB-IoT is deployed as a standalone with at least $180\,kHz$ of the GSM spectrum. All the transmit power at the base station can be used for NB-IoT, which thus significantly enhances the coverage.
\end{itemize}
\begin{figure}[]
        \centering
        \includegraphics[width=0.6\textwidth]{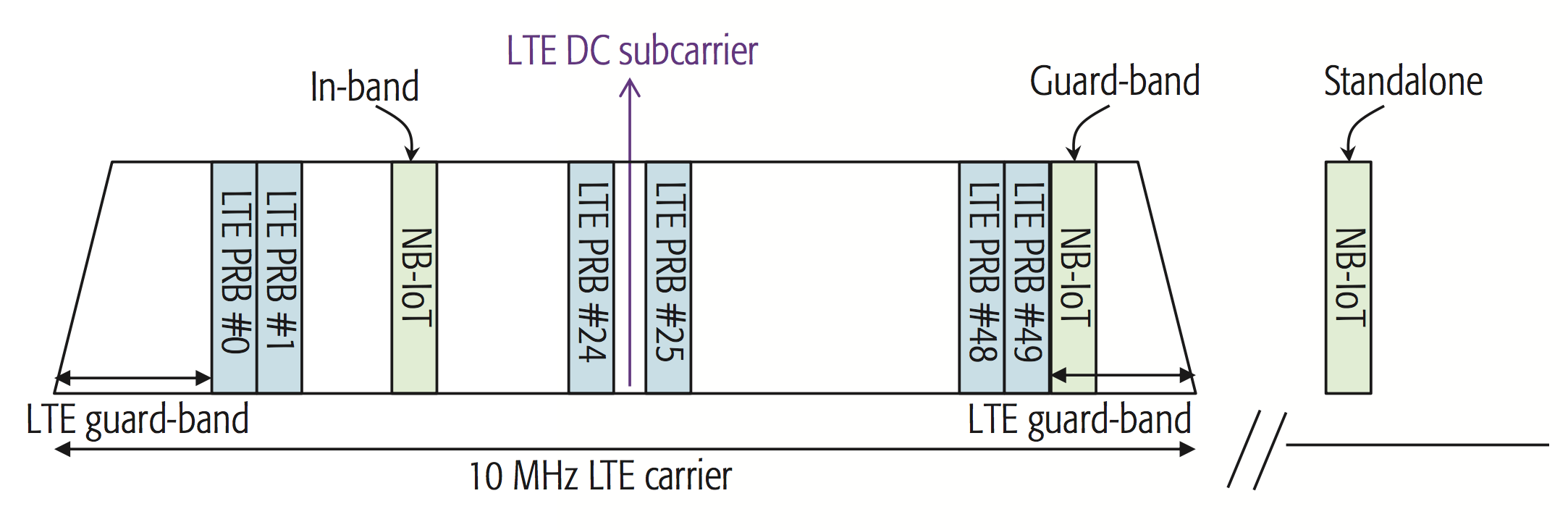}
        \caption{NB-IoT deployment\cite{journal:wang_primer}.}
        \label{fig_bandmodes}
\end{figure}
Similar to existing LTE User Equipments (UEs), an NB-IoT UE is only required to search for a carrier on a $100\,kHz$ raster, referred as anchor carrier\cite{journal:wang_primer}, intended for facilitating UE initial synchronization. An anchor carrier can only be placed in certain PRBs.
NB-IoT reuses the LTE design extensively, including the numerology, with the Orthogonal Frequency Division Multiple Access (OFDMA) for the Downlink, Single Carrier Frequency Division Multiple Access (SC-FDMA) for the Uplink, channel coding, rate matching, interleaving, and so on\cite{journal:wang_primer,3gpp:36211,3gpp:36212,3gpp:36213}.

\subsubsection*{Frequency error constraints for NB-IoT}

Due to the NB-IoT PRB deployment (in \textit{in-band} operation mode), the UE must be able to search the carrier of NB-PRB with a CFO up to $\pm 7.5\,kHz$, according to \cite{conference:ratasuk_overviewNBIoT} and \cite{journal:wang_primer}, during the synchronization procedures. Furthermore the minimum subcarrier spacing of $3.75 \, kHz$ must be taken into account as constraint for Doppler.

        Focussing on the considered scenario, the Doppler experienced by  the i-th user  of the same footprint (in the Downlink channel) and, viceversa, on the satellite with  respect to the i-th user  of the same footprint (in the Uplink channel) can be describer as $f_{di}=f_{d_{common}}+\Delta f_{di}$, where $f_{d_{common}}$ is the common part of Doppler experienced by every user in the same footprint while $\Delta f_{di}$, the differential part, depends on the relative positions of users in the footprint. 

        Regarding the Downlink (DL) user link (which we define as Sat $\rightarrow$ nUE), the differential Doppler is not an issue since each UE has to compensate it own experienced Doppler $f_{di}$  and CFO impairments. In fact, the whole bandwidth of $180\,kHz$ will be received by each UE under the same Doppler condition, with negligible effects on the single subcarriers. 
        
        On the other hand, on the UpLink (UL) user link (which we define as nUE $\rightarrow$ Sat), due to the fact that each nUE generates its own signal and SC-FDMA is used, $\Delta f_{di}$ must be somehow compensated such that the frame structure seen by the satellite does not contain overlapping information among subcarriers. Specifically $\Delta f_{di}$ must be mitigated in case its value is above the LTE Doppler constraints of $950\,Hz$ ,which is derived using 3GPP specification about mobile UEs, a carrier frequency of $2\,GHz$ and a maximum relative speed at $500\,km/h$\cite{3gpp:25913}.
More details can be also found in \cite{3gpp:36304}.

\begin{figure}[]
        \centering
        \includegraphics[width=0.6\textwidth]{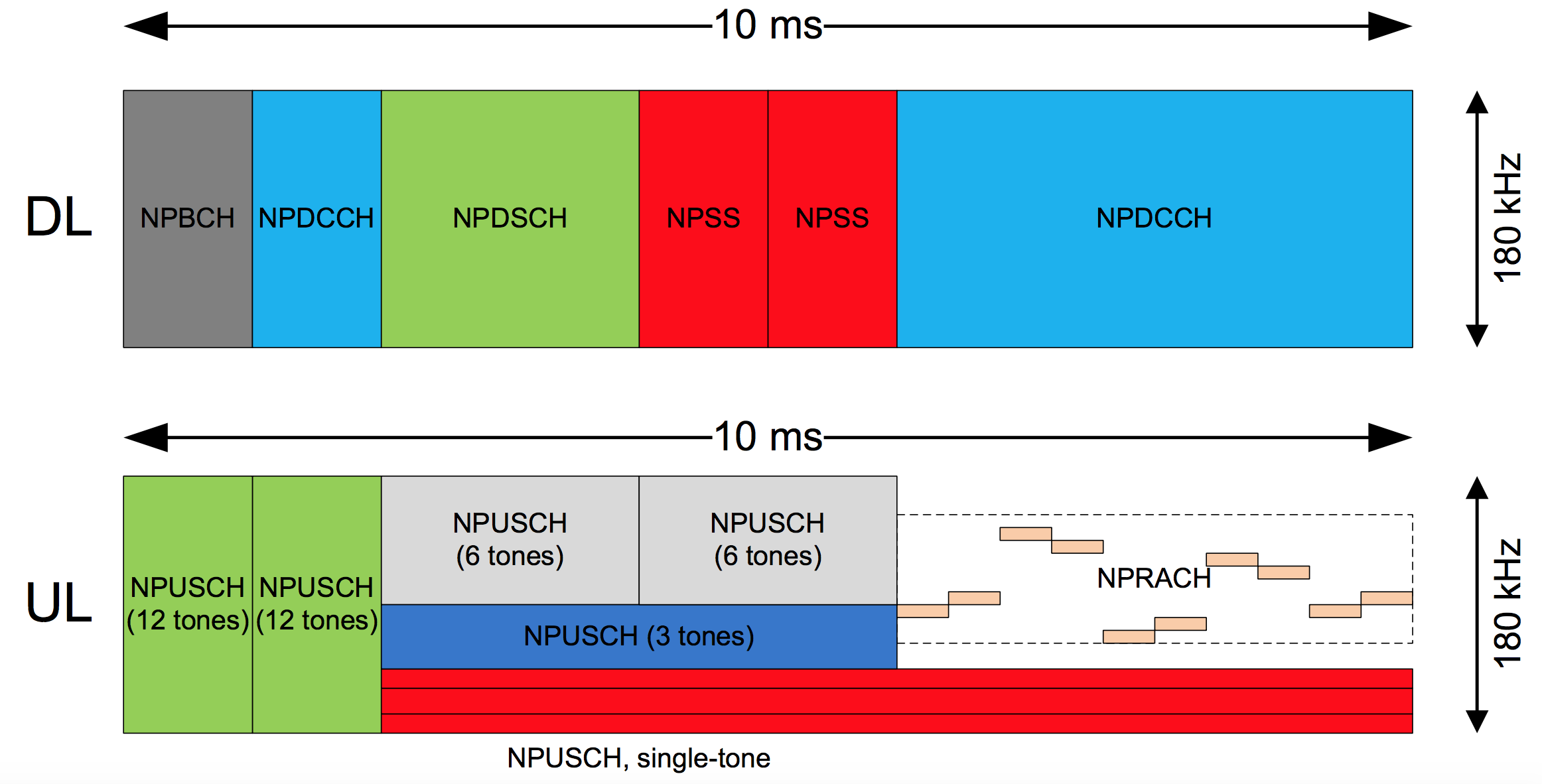}
        \caption{Example NB-IoT design\cite{conference:mangalvedhe_nbiot_2016}}
        \label{fig_DLULch}
\end{figure}

\subsubsection*{Analysis on frequency errors for NB-IoT waveforms}

The aim of this section is to describe and assess the impact of both differential Carrier Frequency Offset (CFO) and differential Doppler among carriers coming from different UEs in the Uplink channel. The parameters used in the below formulation are defined in Table~\ref{tab:NBIOT_Definitions}.        

The satellite system considered is LEO with transparent payload and variable zenith heights from 600 up to 1500 km. A condition case of $90^o$ of elevation angle has been considered, which is the worst case for the problem of the differential Doppler among UEs carriers. The satellite speed is constant and computed as:
\[
\omega_{sat} = \sqrt{\frac{G\cdot M_e}{(R_e+h_{inst})^3}}
\] 
\[
\omega_{sat} = \text{angular velocity} \quad G = \text{gravitational constant}
\]
\[
M_e = \text{Earth mass} \quad R_e = \text{Earth radius} \quad h_{inst} = \text{satellite height}
\]

Source and receiver velocities are much smaller then the propagation velocity of the wave in the medium ($c=3 \cdot 10^8$), hence, the Doppler effect can be assumed symmetric in both channel directions:
\[
f_d \overset{\Delta}{=} \biggl( \frac{c+v_r}{c+v_s} \biggr) f_0 \approx \biggl( 1 + \frac{\Delta v}{c} \biggr) f_0
\]
\[
f_d = \text{doppler shift} \quad v_s= \text{source velocity} \quad v_r= \text{receiver velocity}
\]
\[
\Delta v = |v_r-v_s| \quad f_o = \text{system carrier frequency}
\]
The Doppler variation, assuming a fixed carrier frequency and satellite orbit, is time dependent and it is mainly due to the radial component of the velocity which varies with respect to the relative position between UE and satellite. Regarding the CFO compensation in the Downlink channel, in the following analysis we assume the use of the frequency advance method in \cite{conference:jungnickel_lte_2012,conference:jungnickel_demonstration_2007}. The offset compensation should be intended with respect to the eNB reference frequency which is acquired by the UE at the switch on procedure every time the UE resumes from the RRC\_IDLE state\cite{3gpp:36304}.

In the following, CFO and Doppler effects will be investigated separately in order to highlight the effect given by each component. It is worth mentioning that the index k refers to the k-th user.\begin{itemize}
        \item \textbf{Carrier Frequency Offset:}
        
        The assumption is that Doppler effects are neglected in both eNB $\leftrightarrow$ Sat and Sat $\leftrightarrow$ nUE\textsubscript{k links of the UpLink channel} , while only the Local Oscillator (LO) imperfection of the nUE\textsubscript{k} introduces an offset $f_k$ and both the eNB and satellite LO (in case of frequency conversion) are perfectly centered in $f_0$, which is the carrier frequency of the system.
        
\begin{table}[t!]
\renewcommand{\arraystretch}{1.3}
\caption{Nomenclature for CFO and Doppler analyses.}
\label{tab:NBIOT_Definitions}
\centering
\begin{tabular}{|c|c|}
\hline
\bfseries Definition & \bfseries Symbol \\
\hline
\hline
Number of nUE in the cell & $N$\\
\hline
Received baseband signal nUE\textsubscript{k}$\rightarrow$ eNB & $y_R$\\
\hline
Transmitted baseband signal nUE\textsubscript{k}$\rightarrow$ eNB & $x_{UE_k}$\\
\hline
Carrier Frequency Offset (common) of nUEs w.r.t. eNB & $f_{ko}$\\
\hline 
Differential Frequency Offset of nUEs & $f_{\Delta k}$\\
\hline
Frequency offset added by nUE\textsubscript{k} local oscillator & $f_k=f_{ko}+f_{\Delta k}$\\
\hline
Doppler added at time instant $t$ for the $k$-th user (Sat $\rightarrow$ nUE) & $f_{d_k}(t)$\\
\hline
Doppler added at time instant $t+\tau$ for the $k$-th user (nUE $\rightarrow$ Sat) & $f_{d_k}(t+\tau)$\\
\hline
\end{tabular}
\end{table}

        \textbf{\\ Uplink  (nUEs $\mathbf{\rightarrow}$ eNB) }

        Each k-th  nUE transmits it own signal affected by a frequency offset caused by the mismatch between its own carrier frequency (non perfect LO) and the system carrier frequency (assuming perfect LO for eNB). Due to the differential frequency offset amongst different nUEs, the received baseband signal at the eNB is given by the superposition of each signal as follows:
        \[
        y_{R}=\sum_{k=1}^{N} x_{{UE}_k} e^{-j 2 \pi f_{k} t}=e^{-j 2 \pi f_{ko} t}\sum_{k=1}^{N} x_{{UE}_k} e^{-j 2 \pi f_{\Delta k} t}
        \]
        While the multiple access of baseband signals associated to each UE is managed by means of SC-FDMA, the differential frequency offset amongst nUEs, depending on the amount of differential offset, can be a source of degradation for the system which must be compensated at the nUE through frequency advance due to the slotted structure of SC-FDMA waveform. This procedure is based on the downlink broadcast signal. 
\\
        
\item \textbf{Doppler}\\
        The aim of this paragraph is to repeat the analysis reported above, this time taking into account Doppler effect, in particular by focussing on possible differences on the two cases.
        
The assumption is that the LOs of eNB and nUE\textsubscript{k} are perfectly centered in $f_0$ and no frequency offset due to LOs are introduced. For the sake of simplicity, Doppler effect introduced by the  eNB $\leftrightarrow$ Sat link is neglected, so an ideal link is assumed. This is justified by noticing that, for the  the eNB $\leftrightarrow$ Sat link, the Doppler is applied to the whole composite signal and it does not introduce differential Doppler effects, hence it can be compensated even at the eNB. As before, we define the parameters used:

\begin{table}[t!]
\renewcommand{\arraystretch}{1.3}
\caption{Simulation parameters.}
\label{tab:NBIOT_SimParam}
\centering
\begin{tabular}{|c|c|}
\hline
\bfseries Parameter & \bfseries Value \\
\hline
\hline
Carrier Frequency & $2.2$ GHz\\
\hline
Satellite altitude range & $600-1500$ km\\
\hline
Elevation angle & $90^{\circ}$\\
\hline
Minimum elevation angle & $10^{\circ}$\\
\hline
Reference UEs reciprocal distance & $40-200$ km\\
\hline
\end{tabular}
\end{table}

        \textbf{\\ Uplink only (nUEs $\mathbf{\rightarrow}$ eNB)\\ }
        Since each k-th nUEs is in a different position on the coverage area, it experience a different Doppler value at the same time instant. This means that, differently from the CFO analysis, there is a correlation between the users positions and the differential Doppler experienced. Due this differential Doppler among different nUEs, the received baseband signal at the eNB is given by the superposition of each signal as follows:
        \[
        y_{R}=\sum_{k=1}^{N} x_{{UE}_k} e^{-j 2 \pi f_{d_{k}}(t) t}=e^{-j 2 \pi f_{d_{1}}(t) t}\sum_{k=1}^{N} x_{{UE}_k} e^{-j 2 \pi \Delta f_{d_{k}}(t) t}
        \]

        where $ f_{d_{k}}(t) = f_{d_{1}}(t) + \Delta f_{d_{k}}(t)$ and $f_{d_{1}}$ is the Doppler of one user in the cell
        taken as reference user.

        While the common part (specified by the term $e^{-j 2 \pi f_{d_{1}}(t) t}$) can be compensated at the eNB, the differential Doppler amongst users must be pre-compensated at each $nUE_k$ to avoid dangerous degradations. In order to accomplish this procedure, each nUE must be aware of the instantaneous Doppler generated at the antenna of the satellite and it must somehow mitigate the Doppler effect using a Doppler pre-compensation procedure. It is worth noting that the frequency advance procedure is not an option in this case since, differently from the CFO analysis where the offset is introduced by the UEs, the carrier error is generated by the channel. 
\end{itemize}

\subsubsection*{Differential Doppler assessment}

\begin{figure}[t!]
     \subfloat[Doppler curves of reference UEs, $h_{SAT}=600\,km$.\label{doppler_curve600}]{%
       \includegraphics[width=0.5\textwidth]{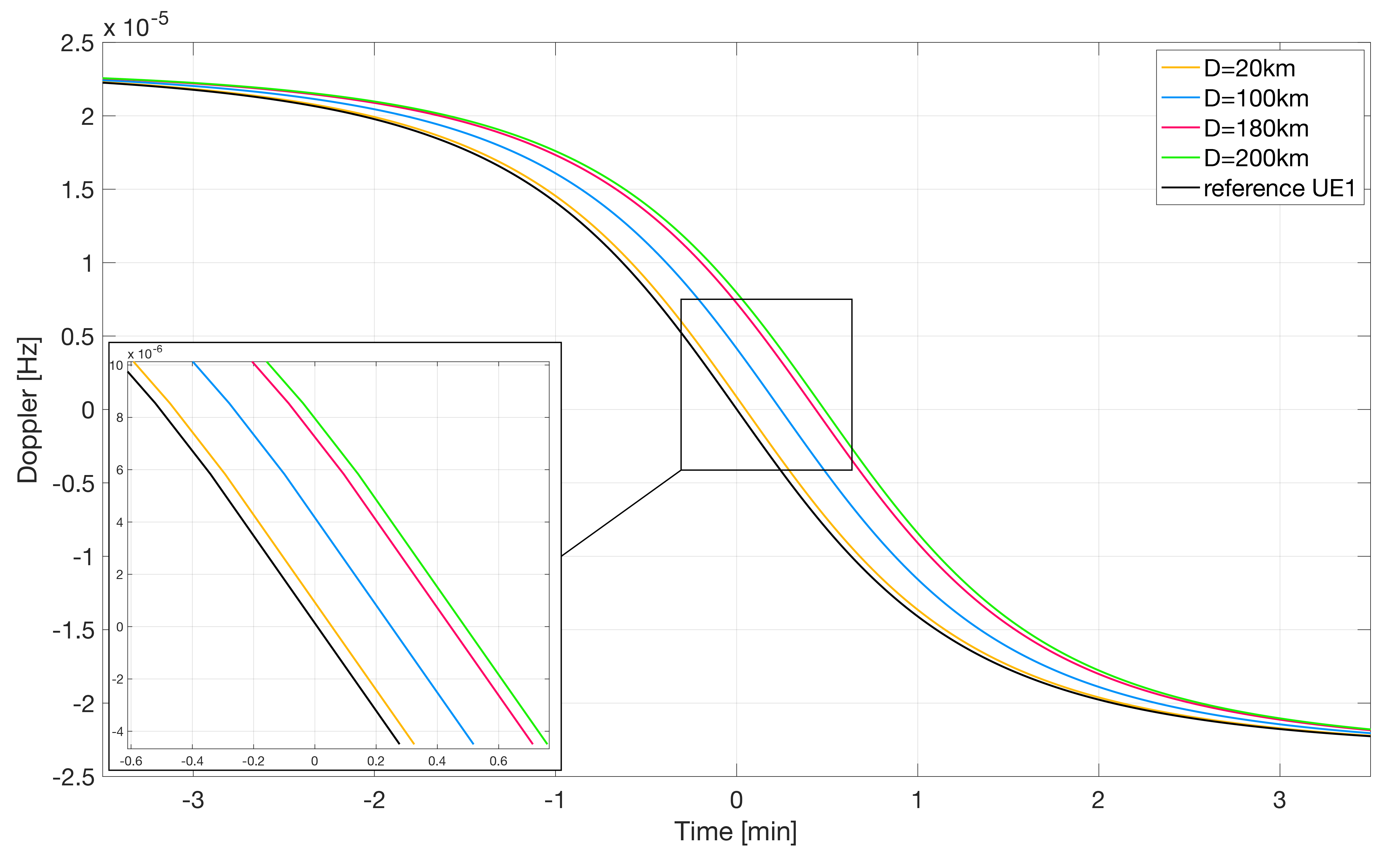}
     }
     \hfill
     \subfloat[Doppler curves of reference UEs, $h_{SAT}=1500\,km$.\label{doppler_curve1500}.\label{fig:ArchitectureA2}]{%
       \includegraphics[width=0.5\textwidth]{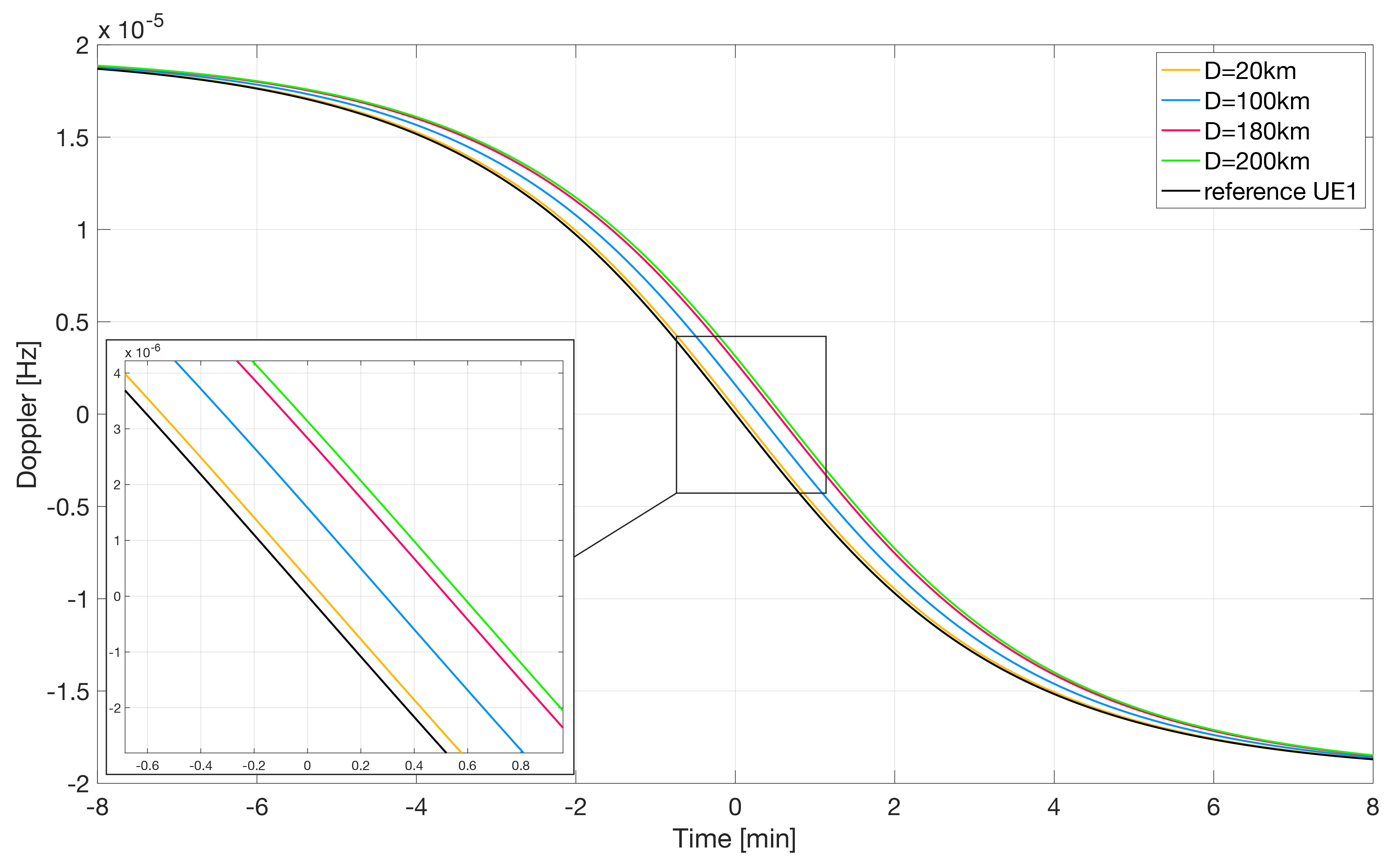}
     }
     \caption{Doppler curves of reference UEs in the NB-IoT scenario.}
     \label{fig_doppler_curves}
\end{figure}

\begin{figure}[h!]
\centering
\includegraphics[width=0.6\textwidth]{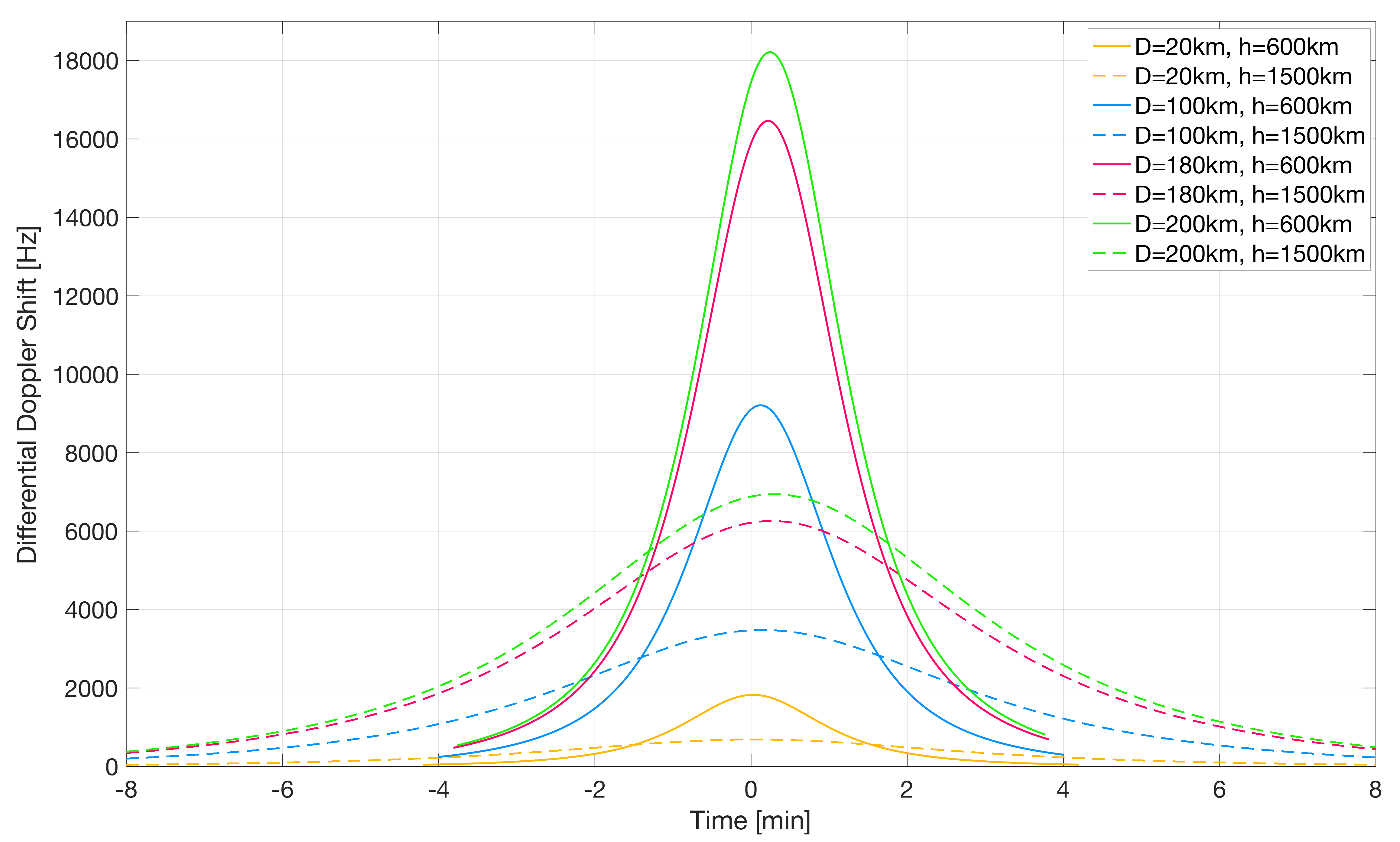}
\caption{Doppler shift analysis for the NB-IoT scenario.}
\label{fig_dopplershift}
\end{figure}

In this paragraph, a preliminary assessment on the evaluation of the differential Doppler over LEO satellites is shown. The assessment has been performed using an analytical characterization of the Doppler, which can be found in \cite{Intro_UniBo} and \cite{journal:ali_doppler_1998}. The purpose is to quantify the differential Doppler in worst case conditions and compare the obtained results with the constraint given by NB-IoT. For the purpose of the analysis, the following parameters are used:


Figure \ref{fig_doppler_curves} shows the Doppler behaviour over time for two UEs placed at different distances in the coverage area. The difference between two curves of the same considered distance is the differential Doppler as seen by the satellite antenna at each timing instant. The differential Doppler is highlighted in Figure~\ref{fig_dopplershift}, in which the changing parameter is the carrier frequency and the satellite altitude. 

As expected, the higher the orbit the lower the differential Doppler. The common Doppler can be compensated by means of a GNSS receiver, according to \cite{Intro_UniBo}. On the other hand, it is worth recalling that the maximum differential Doppler, according to LTE Doppler constraint, should be below 950 Hz.

While the purpose of this section is to highlight the problem assessment only, there are some solutions which can be proposed and which will be part of future works.
The aim of these solutions is to mitigate the level of frequency errors down to a value tolerable by the UE device.

In fact, if the Doppler can be reduced down to the limit reported in the previous section using additional strategies, like for example  position tracking solutions used in \cite{Intro_UniBo}, the use NB-IoT over LEO can be facilitated. While this solution is compatible with the current devices, a GNSS device can drain part of the battery life of the UE.        
        Another technique to be taken into account for frequency/Doppler mitigation is the so called \textit{frequency advance}\cite{conference:jungnickel_lte_2012,conference:jungnickel_demonstration_2007} which allows to estimate the frequency offset in the forward link and then use this information in the reverse link for an effective frequency offset reduction and a consequent improvement in frequency multiplexing in the uplink.
However, frequency advance is not designed for differential Doppler effects and some adaptations may be required.


\subsubsection{PHY/MAC Procedures}
\subsubsection*{Timing Advance}

Timing Advance (TA) is a procedure that is used to control uplink signal transmission timing, in order to maintain a perfect alignment among all the transmission from all the UE served by the same eNB. The TA in NB-IoT is performed from the UE assisted by the eNB and follow the same steps as legacy LTE. Upon reception of a timing advance command, the UE shall adjust uplink transmission timing for NPUSCH based on the received timing advance command. Like in LTE, the timing advance command indicates the change of the uplink timing relative to the current uplink timing as multiples of $16\,T_s$, where $T_s=1/(15000 \times 2048)\,s$ is the time unit for LTE. Transmission of the uplink radio frame from the UE shall start $(N_{TA} + N_{TA_{offset}}) \times T_s$ seconds before the start of the corresponding downlink radio frame at the UE. Where $N_{TA}$ is the timing offset between uplink and downlink radio frames at the UE, expressed in $T_s$ units, and it is define as $N_{TA}=T_A \times 16$. The timing advance command, $T_A$, is given be the eNB to the UE in order to properly perform the time adjustment. If the timing advance command reception ends in the downlink subframe $n$ the uplink transmission timing adjustment should be applied from the first available NB-IoT uplink slot following the end of $n+12$ downlink subframe and the first available NB-IoT uplink slot for NPUSCH transmission. 

This leads to the conclusion that it is possible to compensate a time misalignment among UEs for the uplink transmission, up to a maximum of $0,67\,ms$ fo NB-IoT, which is also the maximum supported value for TA in legacy LTE\cite{3gpp:36211,3gpp:36213,3gpp:36321}.

\begin{figure}[h!]
\centering
\includegraphics[width=0.6\textwidth]{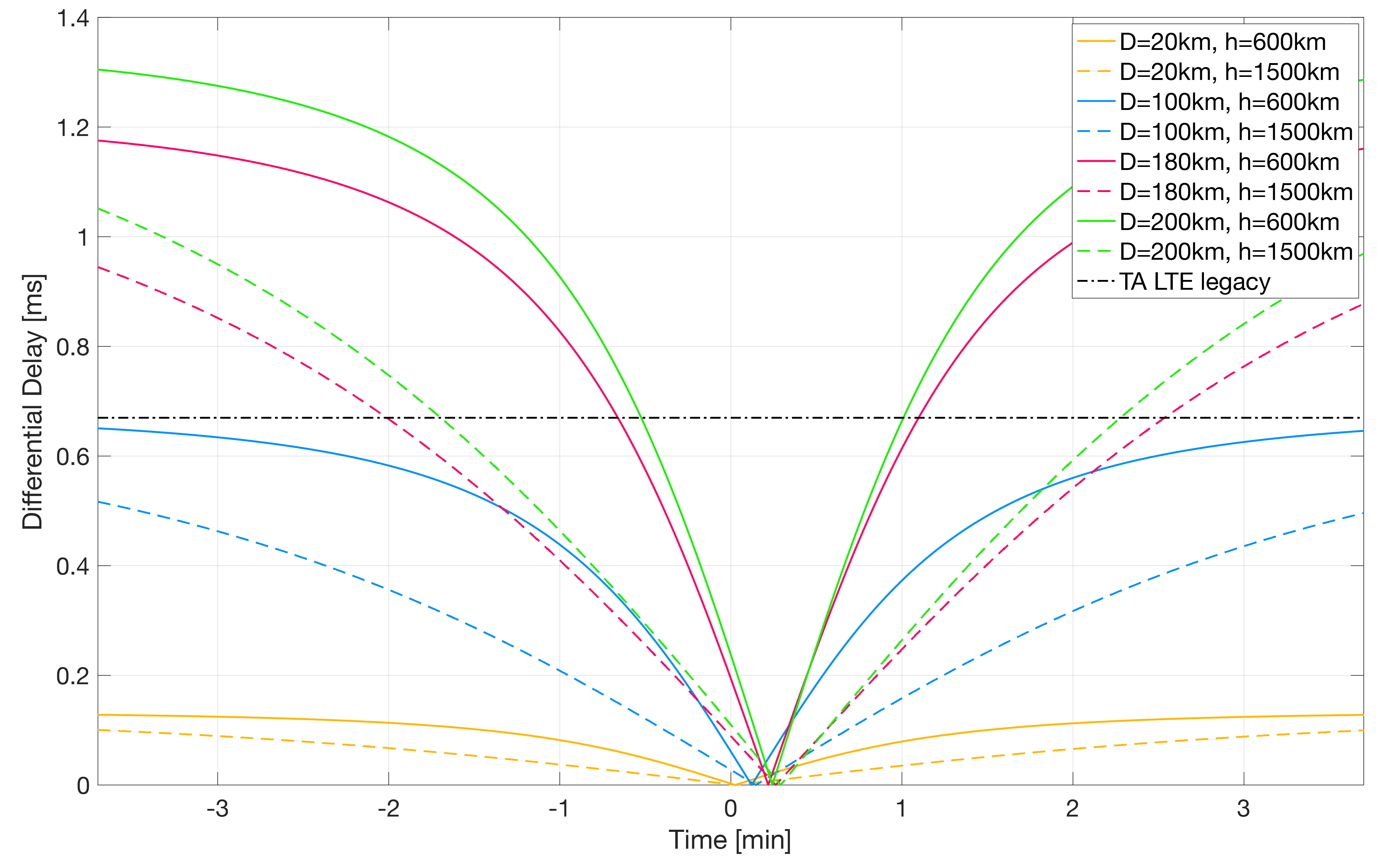}
\caption{Timing Advance analysis for the NB-IoT scenario.}
\label{fig_ TAanalysis}
\end{figure}

In the reference scenario, the maximum allowed TA must correspond to the maximum difference of the travel time of signals between UEs and Sat, i.e. in the worst case the travel time difference of the signals of the users at the edge of the footprint should be under the maximum allowed value for TA. As shown in Figure~\ref{fig_ TAanalysis} accordingly to the Sat altitude and maximum UEs distance inside the footprint, there is a time window such that the worst case differential delay is under the maximum allowed TA threshold. Performing the transmission inside this time window does not require modification to the standard. Adding a time offset which takes into account the propagation delay of the satellite channel, depending on geographical positions of UEs and Sat, could be an alternative solution to overcome the limitation of the TA command.

\subsubsection*{Random Access}
Generally, the RA procedure for NB-IoT follows the same message flow of LTE and NR, then the already mentioned RA procedure in \cite{3GPP_38211, 3GPP_38213, 3GPP_38321} will be taken as reference, and the differences will be highlighted in the following. 

The main difference comes from the repeated transmission of each message in the RA procedure. In order to serve UEs in different coverage classes, whom experience different ranges of path loss, the network can configure up to three Narrowband Physical Random Access Channel (NPRACH) resource configurations in a cell. UE measures its downlink received signal power to estimate which of the three coverage levels it belongs to. In each configuration, a repetition value is specified for repeating the messages in the RA procedure as well as the level of transmission power. Furthermore NB-IoT allows flexible configuration of NPRACH resources in a time-frequency resource grid\cite{3gpp:36211}. 
NB-IoT supports only contention-based random access, unlike LTE which allows both contention-based and contention-free procedures. Referring to the message flow in Figure~\ref{fig:RAprocedure} a UE starts with the transmission of a preamble among one of the 64 available temporary preambles. The NB-IoT preamble consists of 4 symbol groups, with each symbol group comprising one CP and 5 identical symbols, the sequence. Only 2 preamble formats are defined for NB-IoT, which both have a sequence duration of $1.333\,ms$ and differ in their cyclic prefix length, $67\,\mu s$ for "Format 0" and $267\,\mu s$ for "Format 1"\cite{3gpp:36211}. Each symbol group is transmitted on a different $3.75\,kHz$ subcarrier then the tone frequency index changes from one symbol group to another, resulting on a frequency hopping applied on symbol group granularity. The frequency hopping is restricted to 12 contiguous subcarriers but the hopping scheme is different for each repetition of the preamble. Up to 128 repetitions of preamble are allowed for each RA attempt, based on the coverage level of the UE\cite{3gpp:36331}. 
The eNB responds to the request, on the Narrowband Physical Downlink Control Channel (NPDCCH), with a RAR message. The latter should be received by the UE within a RA time window between starting after the transmission of the last preamble repetition plus 4 or 64 subframes, in relation to the repetition number. The maximum duration of the RAR time window is extended, compared to the one of LTE, up to $10.24\,s$\cite{3gpp:36331}. After the counter expiration, the UE can attempt a new RA procedure, up to 10 tentatives.
Then if the maximum number of tentatives is reached without success, the UE proceeds to the next coverage level, if this level is configured. If the total number of access attempts is reached, an associated failure is reported to the RRC. With the RAR, the UE gets a temporary C-RNTI, the timing advance command as well as additional informations, because of the NB-IoT specific uplink transmission scheme. Then a new formula for deriving Random Access Radio Network Temporary Identifier (RA-RNTI) is defined\cite{3gpp:36321,3gpp:36213,conference:ratasuk_overviewNBIoT}. 
Further, the RAR provides the UL grant necessary for transmission of message 3 over the Narrowband Physical Uplink Shared Channel (NPUSCH). Then the remaining procedure is done like in LTE, the UE sends an identification and upon reception of the Contention Resolution indicating that the random access procedure is successfully finalized. Likewise RAR time window, also the MAC contention resolution timer is extended up to $10.24\,s$\cite{3gpp:36101,3gpp:36211,3gpp:36213,3gpp:36304,3gpp:36321,3gpp:36331,conference:ratasuk_overviewNBIoT,journal:wang_primer,whitepaper:RohdeSchwarz,whitepaper:Huawei}.

The extension of RAR time windows, among message 1 and 2, and contention resolution timer, among message 3 and 4 up to $10.24\,s$, allow to cope with the characteristic delay of the satellite channel. Therefore the same RA access procedure for NB-IoT could be used also in the considered satellite scenario, without requiring any modification to the standard. 


\subsubsection*{HARQ}

The Hybrid Automatic Repeat reQuest (HARQ) procedure for NB-IoT is similar to the one in LTE and NR and, therefore, the already mentioned HARQ procedure in \cite{3GPP_38213,3GPP_38321} will be taken as reference, with the differences highlighted in the following.

In order to enable low-complexity UE implementation, NB-IoT allows only one HARQ process, rather then 8 parallel process of LTE, in both downlink and uplink, and allows longer UE decoding time for both Narrowband Physical Downlink Control Channel (NPDCCH) and Narrowband Physical Downlink Shared Channel (NPDSCH). From Release 14, if supported, it is also possible to enable 2 parallel HARQ processes at the MAC entity, in both uplink and downlink, \cite{3gpp:36213}. An asynchronous, adaptive HARQ procedure is adopted to support scheduling flexibility. In order to schedule downlink data or uplink data the eNB conveys a scheduling command through a Downlink Control Indicator (DCI), carried by Narrowband Physical Downlink Control Channel (NPDCCH). DCI could be repeated in order to achieve further coverage enhancement. Repetitions are sent in contiguous suframes and each repetition occupies one subframe. For what concern the downlink, in order to cope with the reduced computing capabilities of NB-IoT devices, the time offset between NPDCCH and the associated NPDSCH is at least $4\,ms$,
instead to schedule the latter in the same TTI as it is for legacy LTE\cite{3gpp:36213}. After receiving NPDSCH, the UE needs to send back a HARQ acknowledgment using NPUSCH Format 2, which is scheduled at least $12\,ms$
after receiving NPDSCH, for the sake of reduced complexity constraints. Similarly, for the uplink the time offset between the end of NPDCCH and the beginning of the associated NPUSCH is at least $8\,ms$\cite{3gpp:36213}.
After completing the NPUSCH transmission, the UE does not expect the reception of the associated HARQ acknowledgment before $3\,ms$\cite{3gpp:36213}.
These relaxed time constraints allow ample decoding time for the UE. The resources to be allocated, as well as the precise time offsets are indicated in the DCI.

In LTE a HARQ process is associated with a Transport Block (TB) in a given TTI. Due to the multiple retransmissions for the coverage enhancement, the HARQ entity in NB-IoT invokes the same HARQ process for each retransmission that is part of the same bundle. Within a bundle the retransmissions are non-adaptive and are triggered without waiting for feedback from previous transmissions according to the maximum number of repetitions established for that coverage level. A downlink assignment or an uplink grant, for downlink and uplink HARQ operations respectively, corresponding to a new transmission or a retransmission of the bundle, is received after the last repetition of the bundle. If a NACK is received and a retransmission is required, then the whole bundle is retransmitted \cite{3gpp:36213,3gpp:36321,3gpp:36331,journal:wang_primer}.

\section{Conclusions}
\label{sec:Conclusions}
In this paper, moving from the architecture and deployment options currently being discussed within 3GPP standardisation activities for Non-Terrestrial Networks, we assessed the impact of typical satellite channel impairments, \emph{i.e.}, large path losses, delays, and Doppler shifts, on the New Radio air interface and PHY/MAC procedures for both mobile broadband and machine type services. In particular, focusing on the enhanced MobileBroadBand scenario, a comparison between the CP-OFDM and f-OFDM waveform has been discussed in terms of OOBE in the presence of TWTA non-linearities, showing that, although the f-OFDM waveform provides benefits from the bandwidth point of view, it also has an increased sensitivity to distortions and larger PAPR. With respect to the PHY and MAC layer procedures, on the one hand the Random Access and Timing Advance do not pose any peculiar issue thanks to the architecture based on the deployment of on-ground Relay Nodes. On the other hand, the HARQ procedure is deeply impacted by the large delays and, in particular, these values might require a significantly larger number of parallel HARQ processes, which also affects the soft buffer sizes. Different solutions are proposed to keep the number of HARQ processes and buffer size under control. As for the NB-IoT service, the system architecture does not include Relay Nodes and, thus, the impact of large Doppler shifts is a limiting factor. To circumvent this challenge, a Frequency Advance procedure has been proposed. With respect to the large delays, been shown that the Random Access and HARQ procedures can operate with no modifications, while the Timing Advance procedure might pose technical challenges only when the information transmission is scheduled outside a specific time window. Future studies will include further analyses on the PHY/MAC procedures, also taking into account additional parameters and algorithms specifications from the standardisation, and the inclusion of non-linear compensation techniques for the f-OFDM waveform.

\ifCLASSOPTIONcompsoc
  \section*{Acknowledgments}
\else
  \section*{Acknowledgment}
\fi

This work has been supported by European Space Agency (ESA) funded activity SatNEx IV CoO2-PART 1 WI1 “5G Access Solutions Assessment for Satellite Scenarios.” Opinions, interpretations, recommendations and conclusions presented in this paper are those of the authors and are not necessarily endorsed by the European Space Agency.

\ifCLASSOPTIONcaptionsoff
  \newpage
\fi

\end{document}